\documentclass{mscs}
\usepackage[utf8]{inputenc}
\usepackage[english]{babel}
\usepackage{hyperref}
\usepackage{graphics}
\usepackage{stmaryrd}
\usepackage{slashbox}
\usepackage{mathpartir}
\usepackage[all,2cell,ps]{xy}
\usepackage{xypic}
\xyoption{curve}
\UseAllTwocells
\SilentMatrices

\newcommand{\strid}[1]{\cpdfinput{#1.ps}}

\usepackage{shortcuts}

\title{Presentation of a Game Semantics for First-Order Propositional Logic}

\hypersetup{
  pdftitle={\csname @title\endcsname},
  pdfauthor={Samuel Mimram},
  unicode=true,
  colorlinks=true,
  linkcolor=black,
  citecolor=black,
  urlcolor=black
}

\author[Samuel Mimram]{Samuel Mimram\thanks{This work has been supported by the
    ANR Invariants alg\'ebriques des syst\`emes informatiques \hbox{(INVAL)}. Physical
    address: \'Equipe PPS, CNRS and Universit\'e Paris~7, 2 place Jussieu, case
    7017, 75251 Paris cedex 05, France. Email address:
    \hbox{\url{smimram@pps.jussieu.fr}}.}}

\newcommand{\FMR}{\mathbf{MRel}}

\newcommand{\Games}{\mathbf{Games}}
\newcommand{\Alg}[2]{\mathbf{Alg}_{#1}^{#2}}
\newcommand{\intset}[1]{\underline{#1}}

\newcommand{\moves}[1]{M_{#1}}

\newcommand{\qforall}[1]{\forall{#1}.}
\newcommand{\qexists}[1]{\exists{#1}.}

\newtheorem{definition}{Definition}
\newtheorem{remark}[definition]{Remark}
\newtheorem{corollary}[definition]{Corollary}
\newtheorem{lemma}[definition]{Lemma}
\newtheorem{property}[definition]{Property}
\newtheorem{proposition}[definition]{Proposition}
\newtheorem{theorem}[definition]{Theorem}

\renewcommand{\paragraph}[1]{\bigskip\noindent\textbf{#1}\;}

\newcommand{\eqth}[1]{\mathfrak{#1}}
\newcommand{\intp}[1]{\llbracket{#1}\rrbracket}
\newcommand{\represents}[1]{\widetilde{#1}}
\newcommand{\card}[1]{\left|#1\right|}
\newcommand{\size}[1]{\left|#1\right|}
\newcommand{\lrule}[1]{\text{(#1)}}
\newcommand{\before}{\varolessthan}

\begin{document}
\maketitle

\begin{abstract}
  Game semantics aim at describing the interactive behaviour of proofs by
  interpreting formulas as games on which proofs induce strategies. In this
  article, we introduce a game semantics for a fragment of first order
  propositional logic. One of the main difficulties that has to be faced when
  constructing such semantics is to make them precise by characterizing
  definable strategies -- that is strategies which actually behave like a
  proof. This characterization is usually done by restricting to the model to
  strategies satisfying subtle combinatory conditions such as innocence, whose
  preservation under composition is often difficult to show. Here, we present an
  original methodology to achieve this task which requires to combine tools from
  game semantics, rewriting theory and categorical algebra. We introduce a
  diagrammatic presentation of definable strategies by the means of generators
  and relations: those strategies can be generated from a finite set of
  ``atomic'' strategies and that the equality between strategies generated in
  such a way admits a finite axiomatization. These generators satisfy laws which
  are a variation of bialgebras laws, thus bridging algebra and denotational
  semantics in a clean and unexpected way.
\end{abstract}

\newpage

\tableofcontents

\newpage

Denotational semantics were introduced to provide useful abstract invariants of
proofs and programs modulo cut-elimination or reduction. In particular, game
semantics, introduced in the nineties, have been very successful in capturing
precisely the interactive behaviour of programs. In these semantics, every type
is interpreted as a \emph{game}, that is as a set of \emph{moves} that can be
played during the game, together with the rules of the game, formalized by a
partial order on the moves of the game indicating the dependencies between the
moves. Every move in these games is to be played by one of the two players,
called \emph{Proponent} and \emph{Opponent}, who should be thought respectively
as the program and its environment. Interactions between these two players are
sequences of moves respecting the partial order of the game, called
\emph{plays}. Every program is characterized by the set of such interactions
that it can have with its environment during an execution and thus defines a
\emph{strategy} reflecting the interactive behaviour of the program inside the
game specified by the type of the program.

In particular, the notion of \emph{pointer game}, introduced by Hyland and
Ong~\cite{hyland-ong:full-abstraction-pcf} and independently by
Nickau~\cite{nickau:hsf}, gave a fully abstract model of PCF -- a simply-typed
$\lambda$-calculus extended with recursion, conditional branching and
arithmetical constants. It has revealed that PCF programs generate strategies
with partial memory, called \emph{innocent} because they react to Opponent moves
according to their own \emph{view} of the play. Thus innocence -- together with
another condition called \emph{well-bracketing} -- is in their setting a
characterization of \emph{definable} strategies, that is strategies which are
the interpretation of a PCF term. This seminal work has lead to an extremely
successful series of semantics: by relaxing in various ways the innocence
constraint on strategies, it became suddenly possible to characterize the
behaviour of PCF programs extended with imperative features like states,
references, etc.

Unfortunately, these constraints are very specific to game semantics and remain
difficult to link with other areas of computer science or algebra. Moreover, the
conditions used to characterize definable strategies are very subtle and
combinatorial and are thus sometimes difficult to work with. In particular,
showing that these conditions are preserved under composition of strategies
usually requires a fairly large amount of work.

\paragraph{Generating instead of restricting.}
In this paper, we introduce a game semantics for a fragment of first-order
propositional logic and describe a monoidal category $\Games$ of games and
strategies in which the proofs can be interpreted. Instead of characterizing
definable strategies of the model by restricting the strategies we consider to
strategies satisfying particular conditions, we show that we can equivalently
use here a kind of converse approach: we explain how to \emph{generate}
definable strategies by giving a \emph{presentation} of those strategies, \ie we
show that a finite set of definable strategies can be used to generate all
definable strategies by composition and tensoring and finitely axiomatize the
equality between strategies obtained this way.

We we mean precisely by a presentation is a generalization of the usual notion
of presentation of a monoid (or a group, \ldots) presentation to monoidal
categories. For example, consider the bicyclic monoid $B$ whose set of elements
is $\mathbb{N}\times\mathbb{N}$ and whose multiplication $*$ is defined by
\[
(m_1,n_1)*(m_2,n_2)=(m_1-n_1+\max(n_1,m_2),n_2-m_2+\max(n_1,m_2))
\]
This monoid admits the presentation $\pangle{\;p,q\;|\;pq=1\;}$, where $p$ and
$q$ are two \emph{generators} and $pq=1$ is an equation between two elements of
the free monoid $M$ on $\{p,q\}$. This means that $B$ is isomorphic to the free
monoid $M$ on two generators $p$ and $q$ quotiented by the smallest congruence
$\equiv$ (\wrt multiplication) such that $pq\equiv 1$, where $1$ is the unit of
the free monoid. More explicitly, the morphism of monoids $\varphi:M\to B$
defined by $\varphi(p)=(1,0)$ and $\varphi(q)=(0,1)$ is surjective and induces
an injective functor from $M/\equiv$ to $B$: two words $w$ and $w'$ have the
same image under $\varphi$ if and only if $w\equiv w'$.

Similarly, we give in this paper a finite set of typed generators from which we
can generate a free monoidal category $\mathcal{G}$ by composing and tensoring
generators. We moreover give a finite set of typed equations between morphisms
of $\mathcal{G}$ and write $\equiv$ for the smallest congruence (\wrt
composition and tensoring) on morphisms. Then we show that the category
$\mathcal{G}/\equiv$, which is the category $\mathcal{G}$ whose morphisms are
quotiented by the congruence $\equiv$, is equivalent to the category $\Games$ of
definable strategies. As a by-product we obtain the fact that the composite of
two definable strategies is well-defined which was not obvious from the
definition we gave.

\paragraph{Strategies as refinements of the game.}
Game semantics has revealed that proofs in logics describe particular strategies
to explore formulas. A formula $A$ is a syntactic tree expressing in which order
its connectives must be introduced in cut-free proofs of $A$: from the root to
leaves. In this sense, it can be seen as the rules of a game whose moves
correspond to introduction rules of connectives in logics. For instance,
consider a formula $A$ of the form
\begin{equation}
  \label{eq:formula-ffe}
  \qforall x P\quad\Rightarrow\quad\qforall y\qexists z Q
\end{equation}
where $P$ and $Q$ are propositional formulas which may contain free
variables. When searching for a proof of $A$, the $\forall y$ connective must be
introduced before the $\exists z$ connective and the $\forall x$ connective can
be introduced independently. The game -- whose moves are first-order connectives
-- associated to this formula is therefore a partial order on the first-order
connectives of the formula which can be depicted as the following diagram (to be
read from the bottom to the top)
\begin{equation}
  \label{eq:game-ffe}
  \xymatrix{
    &\exists z\\
    \forall x&\ar@{-}[u]\forall y\\
  }
\end{equation}
Existential connectives should be thought as Proponent moves (the strategy gives
a witness for which the formula holds) and the universal connectives as Opponent
moves (the strategy receives a term from its environment, for which it has to
show that the formula holds).

Informally, in a first-order propositional logic, the formula
(\ref{eq:formula-ffe}) can have proofs of the three following shapes
\[
\inferrule{
\inferrule{
\inferrule{
\inferrule{\vdots}
{P[t/x]\vdash Q[t'/z]}
}
{P[t/x]\vdash\qexists z Q}
}
{P[t/x]\vdash \qforall y\qexists z Q}
}{\qforall x P\vdash \qforall y\qexists z Q}
\qquad
\inferrule{
\inferrule{
\inferrule{
\inferrule{\vdots}
{P[t/x]\vdash Q[t'/z]}
}
{P[t/x]\vdash\qexists z Q}
}
{\qforall x P\vdash \qexists z Q}
}{\qforall x P\vdash \qforall y \qexists z Q}
\qquad
\inferrule{
\inferrule{
\inferrule{
\inferrule{\vdots}
{P[t/x]\vdash Q[t'/z]}
}
{\qforall x P\vdash Q[t'/z]}
}
{\qforall x P\vdash \qexists z Q}
}{\qforall x P\vdash \qforall y \qexists z Q}
\]
Here $P[t/x]$ denotes the formula $P$ where every occurrence of the free
variable $x$ has been replaced by the term $t$. These proofs introduce the
connectives in the orders depicted respectively below
\[
\xymatrix{
\ar@{-}[d]\forall x\\
\ar@{-}[d]\forall y\\
\exists z\\
}
\qquad
\xymatrix{
\ar@{-}[d]\forall y\\
\ar@{-}[d]\forall x\\
\exists z\\
}
\qquad
\xymatrix{
\ar@{-}[d]\forall y\\
\ar@{-}[d]\exists z\\
\forall x\\
}
\]
It should be noted that they are all refinements of the partial
order~(\ref{eq:game-ffe}) corresponding to the formula, in the sense that they
have more dependencies between moves: \emph{proofs add causal dependencies
  between connectives.}

\medskip

To understand exactly what dependencies which are added by proofs we are
interested in, we shall examine precisely proofs of the formula
\begin{equation}
  \label{eq:formula-ee}
  \qexists x P\quad\Rightarrow\quad\qexists y Q
\end{equation}
which induces the following game
\[
\xymatrix{
  \exists x&\exists y\\
}
\]
By permuting the use of introduction rules, a proof of the formula
(\ref{eq:formula-ee})
\[
\inferrule{
\inferrule{
\inferrule{\vdots}
{P\vdash Q[t/y]}
}
{P\vdash \qexists y Q}
}
{\qexists x P\vdash \qexists y Q}
\]
might be reorganized as the proof
\[
\inferrule{
\inferrule{
\inferrule{\vdots}
{P\vdash Q[t/y]}
}
{\qexists x P\vdash Q[t/y]}
}
{\qexists x P\vdash \qexists y Q}
\]
if and only if the term $t$ used in the introduction rule of the $\exists y$
connective does not have $x$ as free variable. If the variable $x$ is free in
$t$ then the rule introducing $\exists y$ can only be done after the rule
introducing the $\exists x$ connective. This will be reflected by a causal
dependency in the strategy corresponding to the proof, depicted by an oriented
wire:
\[
\strid{dep_ex}
\]

We thus build a monoidal category $\Games$ of games and strategies. Its objects
are \emph{games}, that is total orders on a set whose elements (the
\emph{moves}) are polarized (they are either Proponent or Opponent moves). Its
morphisms $\sigma:A\to B$ between two objects $A$ and $B$ are the partial orders
$\leq_\sigma$ on the moves of $A$ (with polarities inverted) and $B$ which are
compatible with both the partial orders of $A$ and of $B$, \ie does not create
cycle with those partial orders.

The logic we have chosen to model here (the fragment of first-order
propositional logic without connectives) is deliberately very simple in order to
simplify our presentation of the category $\Games$. We believe however that the
techniques used here are very general and could extend to more expressive
logics.

\section{Presentations of monoidal categories}
\subsection{Monoidal categories}
A \emph{monoidal category} $(\mathcal{C},\otimes,I)$ is a category $\mathcal{C}$
together with a functor
\[
\otimes:\mathcal{C}\times\mathcal{C}\to\mathcal{C}
\]
and natural isomorphisms
\[
\alpha_{A,B,C}:(A\otimes B)\otimes C\to A\otimes(B\otimes C)
\tcomma\quad
\lambda_A:I\otimes A\to A
\qtand
\rho_A:A\otimes I\to A
\]
satisfying coherence axioms~\cite{maclane:cwm}. A symmetric monoidal category
$\mathcal{C}$ is a monoidal category $\mathcal{C}$ together with a natural
isomorphism
\[
\gamma_{A,B}:A\otimes B\to B\otimes A
\]
satisfying coherence axioms and such that
$\gamma_{B,A}\circ\gamma_{A,B}=\id_{A\otimes B}$. A monoidal category
$\mathcal{C}$ is strictly monoidal when the natural isomorphisms $\alpha$,
$\lambda$ and $\rho$ are identities. To simplify our presentation, in the rest
of this paper we only consider strict monoidal categories.  Formally, it can be
shown that it is not restrictive, using MacLane's coherence
theorem~\cite{maclane:cwm}: every monoidal category is monoidally equivalent to
a strict one.

A (strict) \emph{monoidal functor} $F:\mathcal{C}\to\mathcal{D}$ between two
strict monoidal categories $\mathcal{C}$ and $\mathcal{D}$ if a functor $F$
between the underlying categories $\mathcal{C}$ and $\mathcal{D}$ such that
$F(A\otimes B)=F(A)\otimes F(B)$ for every objects $A$ and $B$ of $\mathcal{C}$,
and $F(I)=I$. A monoidal functor $F$ between two strict symmetric monoidal
categories $\mathcal{C}$ and $\mathcal{D}$ is \emph{symmetric} when it
transports the symmetry of $\mathcal{C}$ to the symmetry of $\mathcal{D}$, that
is when $F(\gamma_A)=\gamma_{F(A)}$.

A \emph{monoidal natural transformation} $\theta:F\to G$ between two monoidal
functors $F,G:\mathcal{C}\to\mathcal{D}$ is a natural transformation between the
underlying functors $F$ and $G$ such that $\theta_{A\otimes
  B}=\theta_A\otimes\theta_B$ for every objects $A$ and $B$ of $\mathcal{C}$,
and $\theta_I=\id_I$. A monoidal natural transformation $\theta:F\to G$ between
two strict symmetric monoidal functors is said to be \emph{symmetric}.

Two monoidal categories $\mathcal{C}$ and $\mathcal{D}$ are \emph{monoidally
  equivalent} when there exists a pair of monoidal functors
$F:\mathcal{C}\to\mathcal{D}$ and $G:\mathcal{D}\to\mathcal{C}$ and two
invertible monoidal natural transformations $\eta:\mathrm{Id}_\mathcal{C}\to GF$
and $\varepsilon:FG\to\mathrm{Id}_\mathcal{D}$.

\subsection{Monoidal theories}
A \emph{monoidal theory} $\mathcal{T}$ is a strict monoidal category whose
objects are the natural integers such that the tensor product on objects is
given by addition of integers. By an integer $n$, we mean here the finite
ordinal $\intset{n}=\{0,1,\ldots,n-1\}$ and the addition is given by
$\intset{m}+\intset{n}=\intset{m+n}$. A \emph{symmetric monoidal theory} is a
monoidal theory where the category is moreover required to be symmetric. An
algebra $F$ of a monoidal theory $\mathcal{T}$ in a strict monoidal category
$\mathcal{C}$ is a strict monoidal functor from $\mathcal{T}$ to
$\mathcal{C}$. Every monoidal theory $\mathcal{T}$ and strict monoidal category
$\mathcal{C}$ give rise to a category $\Alg{\mathcal{T}}{\mathcal{C}}$ of
algebras of $\mathcal{T}$ in $\mathcal{C}$ and monoidal natural transformations
between them. Examples of such categories are given in
Section~\ref{section:presentation-rel}. Monoidal theories and symmetric monoidal
theories are sometimes called respectively PRO and PROP, these terms were
introduced by Mac Lane in~\cite{maclane:ca} as abbreviations for respectively
``category with products'' and ``category with products and permutations''.

Monoidal theories generalize equational theories: in this setting, operations
are typed, and can moreover have multiple outputs as well as multiple inputs.

\subsection{Presentations of monoidal categories}
\label{subsection:moncat-presentation}
In this section, we recall the notion of \emph{presentation} of a monoidal
category by the means of 2-dimensional generators and relations.

Suppose that we are given a set $E_1$ whose elements are called \emph{atomic
  types}. We write $E_1^*$ for the free monoid on the set $E_1$ and $i_1:E_1\to
E_1^*$ for the corresponding injection; the product of this monoid is written
$\otimes$ and its unit is written $I$. The elements of $E_1^*$ are called
\emph{types}. Suppose moreover that we are given a set $E_2$, whose elements are
called \emph{generators}, together with two functions $s_1,t_1:E_2\to E_1^*$
which to every generator associate a type called respectively its \emph{source}
and \emph{target}. We call a \emph{signature} such a 4-uple $(E_1,s_1,t_1,E_2)$:
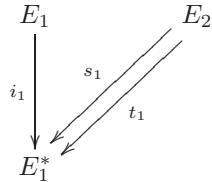
\[
\xymatrix@C=10ex@R=10ex{
  E_1\ar[d]_{i_1}&\ar@<-0.7ex>[dl]_{s_1}\ar@<0.7ex>[dl]^{t_1}E_2\\
  E_1^*&\\
}
\]


\noindent
In particular, every strict monoidal category $\mathcal{C}$ generates a
signature by taking $E_1$ to be the objects of the category $\mathcal{C}$, $E_2$
its morphisms, such that for every morphism $f:A\to B$, we have $s_1(f)=i_1(A)$
and $t_1(f)=i_1(B)$. Conversely, every signature $(E_1,s_1,t_1,E_2)$ generates
a free strict monoidal category $\mathcal{E}$ described as follows. If we write
$E_2^*$ for the morphisms of this category and $i_2:E_2\to E_2^*$ for the
injection of the generators into this category, we get a diagram
\[
\xymatrix@C=10ex@R=10ex{
  E_1\ar[d]_{i_1}&\ar@<-0.7ex>[dl]_{s_1}\ar@<0.7ex>[dl]^{t_1}E_2\ar[d]_{i_2}\\
  E_1^*&\ar@<-0.7ex>[l]_{\overline{s_1}}\ar@<0.7ex>[l]^{\overline{t_1}}E_2^*\\
}
\]
in $\Set$ together with a structure of monoidal category on the graph
\[
\xymatrix@C=10ex@R=10ex{
  E_1^*&\ar@<-0.7ex>[l]_{\overline{s_1}}\ar@<0.7ex>[l]^{\overline{t_1}}E_2^*\\
}
\]
where the morphisms $\overline{s_1},\overline{t_1}:E_2^*\to E_1^*$ are the
morphisms (unique by universality of $E_2^*$) such that $s_1=\overline{s_1}\circ
i_2$ and $t_1=\overline{t_1}\circ i_2$. More explicitly, the category
$\mathcal{E}$ has $E_1^*$ as objects and its set $E_2^*$ of morphisms is the
smallest set such that
\begin{enumerate}
\item there is a morphism $f:A\to B$ in $E_2^*$ for every element $f$ of $E_2$
  such that $s_1(f)=A$ and $t_1(f)=B$ (this is the image by $i_2$ of $f$),
\item there is a morphism $\id_A:A\to A$ in $E_2^*$ for every element $A$ of
  $E_1^*$,
\item for every morphisms $f:A\to B$ and $g:B\to C$ in $E_2^*$ there is a
  morphism $g\circ f:A\to C$ in $E_2^*$,
\item for every morphisms $f:A\to B$ and $g:C\to D$ in $E_2^*$ there is a
  morphism $f\otimes g:A\otimes C\to B\otimes D$ in $E_2^*$,
\end{enumerate}
quotiented by equalities imposing that
\begin{enumerate}
\item composition is associative and admits identities as neutral element,
\item the tensor product is associative and admits $\id_I$ as neutral element,
\item identities form a monoidal natural transformation
  $\id:\mathrm{Id}\to\mathrm{Id}$: for every objects $A$ and $B$,
  \[
  \id_A\otimes\id_B\qeq \id_{A\otimes B}
  \]
\item tensor product and composition are compatible in the sense that for every
  morphisms $f:A\to B$, $g:B\to C$, $f':A'\to B'$ and $g':B'\to C'$,
  \[
  (g\circ f)\otimes(g'\circ f')
  \qeq
  (g\otimes g')\circ (f\otimes f')
  \]
\end{enumerate}
The \emph{size} $\size{f}$ of a morphism $f:A\to B$ in $\mathcal{E}$ is defined
inductively by
\[
\begin{array}{c}
  \size{\id}=0
  \qquad
  \size{f}=1
  \text{ if $f$ is a generator}
  \\
  \size{f_1\otimes f_2}
  =
  \size{f_1}+\size{f_2}
  \qquad
  \size{f_2\circ f_1}
  =
  \size{f_1}+\size{f_2}
\end{array}
\]
In particular, a morphism is of size $0$ if and only if it is an identity.

This construction is a particular case of Street's
2-computads~\cite{street:limit-indexed-by-functors} and Burroni's
polygraphs~\cite{burroni:higher-word} who made precise the sense in which the
generated monoidal category is free on the signature. In particular, the
following notion of equational theory is a specialization of the definition of a
3-polygraph to the case where there is only one 0-cell.


\begin{definition}
  A \textbf{monoidal equational theory} is a 7-uple
  \[
  \eqth{E}=(E_1,s_1,t_1,E_2,s_2,t_2,E_3)
  \]
  where $(E_1,s_1,t_1,E_2)$ is a signature together with a set $E_3$ of
  \emph{equations} and two morphisms $s_2,t_2:E_3\to E_2^*$, as pictured in the
  diagram
  \[
  \xymatrix@C=10ex@R=10ex{
    E_1\ar[d]_{i_1}&\ar@<-0.7ex>[dl]_{s_1}\ar@<0.7ex>[dl]^{t_1}E_2\ar[d]_{i_2}&\ar@<-0.7ex>[dl]_{s_2}\ar@<0.7ex>[dl]^{t_2}E_3\\
    E_1^*&\ar@<-0.7ex>[l]_{\overline{s_1}}\ar@<0.7ex>[l]^{\overline{t_1}}E_2^*\\
  }
  \]
  such that
  \[
  \overline{s_1}\circ s_2=\overline{s_1}\circ t_2
  \qtand
  \overline{t_1}\circ s_2=\overline{t_1}\circ t_2
  \tdot
  \]
\end{definition}
Every equational theory defines a monoidal category $\mathcal{E}/\equiv$
obtained from the monoidal category $\mathcal{E}$ generated by the signature
$(E_1,s_1,t_1,E_2)$ by quotienting the morphisms by the congruence $\equiv$
generated by the equations of the equational theory $\eqth{E}$: it is the
smallest congruence (\wrt both composition and tensoring) such that
$s_2(e)\equiv t_2(e)$ for every element $e\in E_3$. We say that a monoidal
equational theory $\eqth{E}$ is a \emph{presentation} of a strict monoidal
category $\mathcal{M}$ when $\mathcal{M}$ is monoidally equivalent to the
category $\mathcal{E}$ generated by $\eqth{E}$.

We sometimes informally say that an equational theory
\[
\eqth{E}=(E_1,s_1,t_1,E_2,s_2,t_2,E_3)
\]
has a \emph{generator}
\[
f\qcolon A\to B
\]
to mean that $f$ is an element of $E_2$ such that $s_1(f)=A$ and $t_1(f)=B$. We
also say that the equational theory has an \emph{equation}
\[
f\qeq g
\]
to mean that there exists an element $e$ of $E_2$ such that $s_2(e)=f$ and
$t_2(e)=g$.


We say that two equational theories are \emph{equivalent} when they generate
monoidally equivalent categories. A generator $f$ in an equational theory
$\eqth{E}$ is \emph{superfluous} when the equational theory $\eqth{E'}$ obtained
from $\eqth{E}$ by removing the generator $f$ and all equations involving $f$,
is equivalent to $\eqth{E}$. Similarly, an equation $e$ is \emph{superfluous}
when the equational theory $\eqth{E'}$ obtained from $\eqth{E}$ by removing the
equation $e$ is equivalent to $\eqth{E}$. An equational theory is \emph{minimal}
when it does not contain any superfluous generator or equation.

\begin{remark}
  An equational presentation $(E_1,s_1,t_1,E_2,s_2,t_2,E_3)$ where $E_1$ is
  reduced to a set with only one object $\{1\}$ generates a monoidal category
  which is a monoidal theory.
\end{remark}

\subsection{Presented categories as models}
Suppose that a strict monoidal category $\mathcal{M}$ is presented by an
equational theory $\eqth{E}$. We write $\mathcal{E}/\equiv$ for the category
generated by $\eqth{E}$. The proof that $\eqth{E}$ presents $\mathcal{M}$ can
generally be decomposed in three parts:
\begin{enumerate}
\item \emph{$\mathcal{M}$ is a model of the equational theory $\eqth{E}$:} there
  exists a functor $\represents{-}$ from the category $\mathcal{E}/\equiv$ to
  $\mathcal{M}$. This amounts to check that there exists a functor
  $F:\mathcal{E}\to\mathcal{M}$ such that for every morphisms
  $f,g:A\to B$ in $\mathcal{E}$, $f\equiv g$ implies
  $Ff=Fg$.
\item \emph{$\mathcal{M}$ is a fully-complete model of the equational theory
    $\eqth{E}$:} the functor $\represents{-}$ is full.
\item \emph{$\mathcal{M}$ is the initial model of the equational theory
    $\eqth{E}$:} the functor $\represents{-}$ is faithful.
\end{enumerate}
We say that a morphism $f:A\to B$ of $\mathcal{E}/\equiv$ \emph{represents} the
morphism $\represents{f}:\represents{A}\to\represents{B}$ of $\mathcal{M}$.

Usually, the first point is a straightforward verification and the second point
is easy to show. Proving that the functor $\represents{-}$ is faithful often
requires more work. In this paper, we use the methodology introduced by Lafont
in~\cite{lafont:boolean-circuits}. We first define \emph{canonical forms} which
are (not necessarily unique) canonical representatives of the equivalence
classes of morphisms of $\mathcal{E}$ under the congruence $\equiv$ generated by
the equations of $\eqth{E}$ -- proving that every morphism is equal to a
canonical form can be done by induction on the size of the morphisms. Then we
show that the functor $\represents{-}$ is faithful by showing that all the
canonical forms which have the same image under $\represents{-}$ are equal.

It should be noted that this is not the only technique to prove that an
equational theory presents a monoidal category. In particular, Joyal and Street
have used topological methods~\cite{joyal-street:geometry-tensor-calculus} by
giving a geometrical construction of the category generated by a signature, in
which morphisms are equivalence classes under continuous deformation of
progressive plane diagrams (their construction is detailed a bit more in
Section~\ref{subsection:string-diagrams}). Their work is for example extended by
Baez and Langford in~\cite{baez-langford:two-tangles} to give a presentation of
the 2-category of 2-tangles in 4 dimensions. The other general methodology the
author is aware of, is given by Lack in~\cite{lack:composing-props}, by
constructing elaborate monoidal theories from simpler monoidal theories. Namely,
a monoidal theory can be seen as a monad in a particular span category and
monoidal theories can therefore be composed, given a distributive law between
their corresponding monads. We chose not to use those methods because, even
though they can be very helpful to build intuitions, they are difficult to
formalize and even more to mechanize -- we believe indeed that some of the
tedious proofs given in this paper could be somewhat automated.

\subsection{String diagrams}
\label{subsection:string-diagrams}
\emph{String diagrams} provide a convenient way to represent the morphisms in
the category generated by a presentation. Given an object $M$ in a category
$\mathcal{C}$, a morphism $\mu:M\otimes M\to M$ can be drawn graphically as a
device with two inputs and one output of type $M$ as follows:
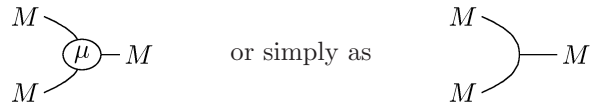
\[
\strid{mult_m_label}
\qquad\text{or simply as}\qquad
\strid{mult_m}
\]
when it is clear from the context which morphism of type $M\otimes M\to M$ we
are picturing (we sometimes even omit the source and target of the
morphisms). Similarly, the identity $\id_M:M\to M$ can be pictured as
\[
\strid{id_m}
\]
The tensor $f\otimes g$ of two morphisms $f:A\to B$ and $g:C\to D$ is obtained
by putting the diagram corresponding to $f$ above the diagram corresponding to
$g$:
\[
\strid{f_x_g}
\]
So, for instance, the morphism $\mu\otimes M:M\otimes M\otimes M\to M\otimes M$
can be drawn diagrammatically as
\[
\strid{mult_x_id_m}
\]
Finally, the composite $g\circ f$ of two morphisms $f:A\to B$ and $g:B\to C$ can
be drawn diagrammatically by putting the diagram corresponding to $g$ at the
right of the diagram corresponding to $f$ and by ``linking the wires''.
\[
\strid{f_o_g}
\]
Thus, the diagram corresponding to the morphism $\mu\circ(\mu\otimes M):M\otimes
M\to M$ is
\[
\strid{mult_assoc_l_m}
\]
The associativity law for monoids (see Section~\ref{subsection:monoids})
\[
\mu\circ(\mu\otimes M)
\qeq
\mu\circ(M\otimes\mu)
\]
can therefore be represented graphically as
\[
\strid{mult_assoc_l_m}
\qeq
\strid{mult_assoc_r_m}
\]

\medskip

Suppose that $(E_1,s_1,t_1,E_2)$ is a signature. Every element $f$ of $E_2$ such
that
\[
s_1(f)=A_1\otimes\cdots\otimes A_m
\qtand
t_1(f)=B_1\otimes\cdots\otimes B_n
\]
where the $A_i$ and $B_i$ are elements of $E_1$, can be represented by a diagram
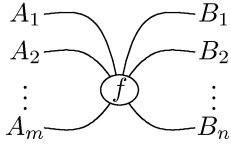
\[
\strid{signature_f}
\]
Bigger diagrams can be constructed from these diagrams by composing and
tensoring them, as explained above. Joyal and Street have shown in details
in~\cite{joyal-street:geometry-tensor-calculus} that the category of those
diagrams, modulo continuous deformations, is precisely the free category
generated by a signature (which they call a tensor scheme). For example, the
equality
\[
(M\otimes\mu)\circ(\mu\otimes M\otimes M)
\qeq
(\mu\otimes M\otimes M)\circ(M\otimes\mu)
\]
in the category $\mathcal{C}$ given in the example above; this can be shown by
continuously deforming the diagram on the left-hand side below into the diagram
on the right-hand side:
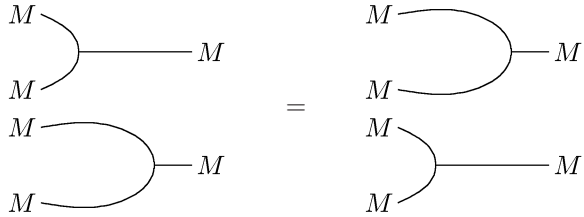
\[
\strid{mu_x_mu_r}
\qeq
\strid{mu_x_mu_l}
\]
All the equalities, given in Section~\ref{subsection:moncat-presentation},
satisfied by the monoidal category generated by a signature have a similar
geometrical interpretation.

\section{Some algebraic structures}
In this section, we recall the categorical formulation of some well-known
algebraic structures (monoids, bialgebras, \ldots). It should be noted that we
give those definitions in the setting of a monoidal category which is \emph{not}
required to be symmetric. We suppose that $(\mathcal{C},\otimes,I)$ is a strict
monoidal category, fixed throughout the section.

\subsection{Symmetric objects}
A \emph{symmetric object} of $\mathcal{C}$ is an object $S$ together with a
morphism
\[
\gamma:S\otimes S\to S\otimes S
\]
called \emph{symmetry} and pictured as
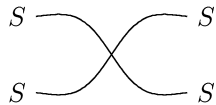
\[
\strid{sym_s}
\]
such that the diagrams
\[
\xymatrix{
  S\otimes S\otimes S\ar[d]_{S\otimes\gamma}\ar[r]^{\gamma\otimes S}&S\otimes S\otimes S\ar[r]^{S\otimes\gamma}&S\otimes S\otimes S\ar[d]^{\gamma\otimes S}\\
  S\otimes S\otimes S\ar[r]_{\gamma\otimes S}&S\otimes S\otimes S\ar[r]_{S\otimes\gamma}&S\otimes S\otimes S\\
}
\qtand
\xymatrix{
  &S\otimes S\ar[dr]^{\gamma}&\\
  S\otimes S\ar[ur]^{\gamma}\ar[rr]_{S\otimes S}&&S\otimes S\\
}
\]
commute. Graphically,
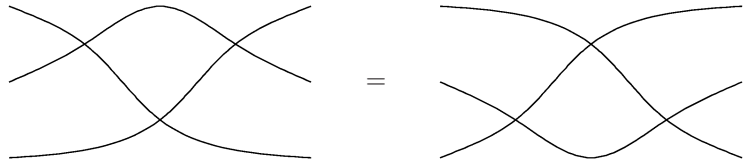
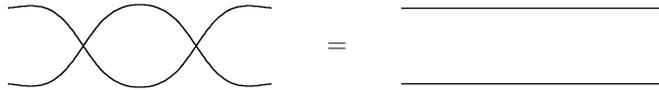
\[
\strid{yang_baxter_r}
\qeq
\strid{yang_baxter_l}
\]
and
\[
\strid{sym_sym}
\qeq
\strid{id_x_id}
\]
These equations are called the Yang-Baxter equations.

\begin{remark}
  When the monoidal category $\mathcal{C}$ is symmetric, every object $S$ has a
  symmetry $\gamma=\gamma_{S,S}$ induced by the symmetry of the category.
\end{remark}

\subsection{Monoids}
\label{subsection:monoids}
A \emph{monoid} $(M,\mu,\eta)$ in $\mathcal{C}$ is an object $M$ together with
two morphisms
\[
\mu : M\otimes M\to M
\qtand
\eta : I\to M
\]
called respectively \emph{multiplication} and \emph{unit} and pictured respectively as
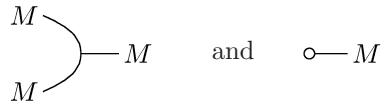
\[
\strid{mult_m}
\qtand
\strid{unit_m}
\]
such that the diagrams
\[
\vxym{
  M\otimes M\otimes M\ar[d]_{M\otimes\mu}\ar[r]^-{\mu\otimes M}&M\otimes M\ar[d]^{\mu}\\
  M\otimes M\ar[r]_{\mu}&M
}
\qtand
\vxym{
  \ar[dr]_{M}I\otimes M\ar[r]^{\eta\otimes M}&M\otimes M\ar[d]_{\mu}&\ar[l]_{M\otimes\eta}M\otimes I\ar[dl]^{M}\\
  &M&
}
\]
commute. Graphically,
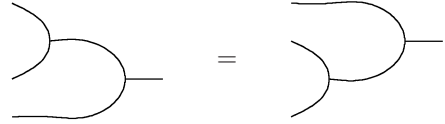
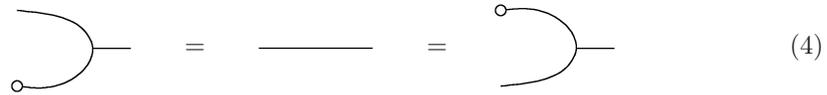
\[
\strid{mult_assoc_l}
\qeq
\strid{mult_assoc_r}
\]
and
\begin{equation}
  \label{eq:monoid_unit}
  \strid{mult_unit_l}
  \qeq
  \strid{mult_unit_c}
  \qeq
  \strid{mult_unit_r}
\end{equation}

A \emph{symmetric monoid} is a monoid which admits a symmetry $\gamma:M\otimes
M\to M\otimes M$ which is compatible with the operations of the monoid in the
sense that it makes the diagrams
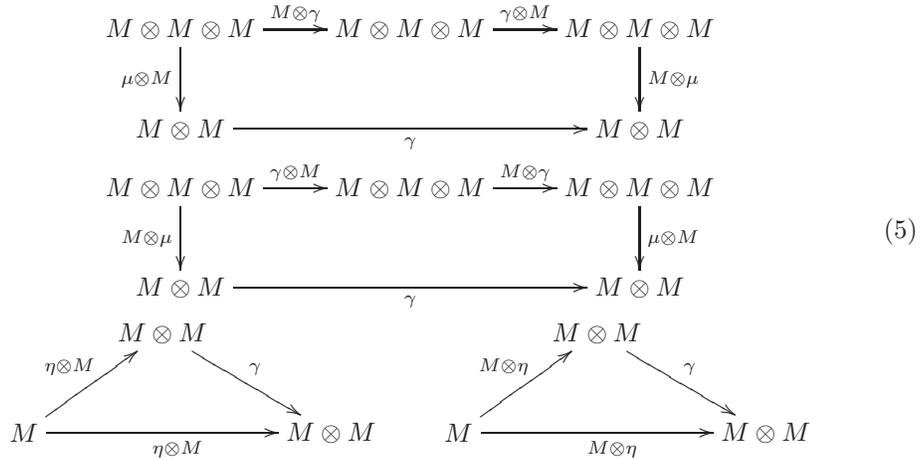
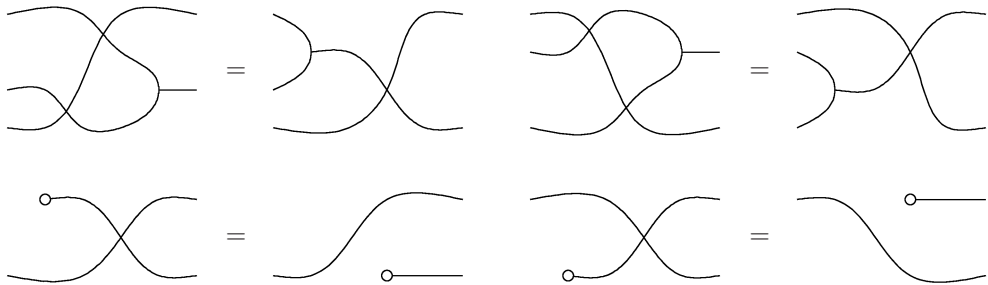
\begin{equation}
  \label{eq:monoid-nat}
  \begin{array}{c}
    \xymatrix{
      \ar[d]_{\mu\otimes M}M\otimes M\otimes M\ar[r]^{M\otimes \gamma}&M\otimes M\otimes M\ar[r]^{\gamma\otimes M}&M\otimes M\otimes M\ar[d]^{M\otimes\mu}\\
      M\otimes M\ar[rr]_{\gamma}&&M\otimes M\\
    }
    \\[4ex]
    \xymatrix{
      \ar[d]_{M\otimes \mu}M\otimes M\otimes M\ar[r]^{\gamma\otimes M}&M\otimes M\otimes M\ar[r]^{M\otimes\gamma}&M\otimes M\otimes M\ar[d]^{\mu\otimes M}\\
      M\otimes M\ar[rr]_{\gamma}&&M\otimes M\\
    }
    \\[4ex]
    \xymatrix{
      &M\otimes M\ar[dr]^{\gamma}&\\
      M\ar[ur]^{\eta\otimes M}\ar[rr]_{\eta\otimes M}&&M\otimes M\\
    }
    \qquad
    \xymatrix{
      &M\otimes M\ar[dr]^{\gamma}&\\
      M\ar[ur]^{M\otimes\eta}\ar[rr]_{M\otimes\eta}&&M\otimes M\\
    }
    \\
  \end{array}
\end{equation}
commute. Graphically,
\[
\begin{array}{cc}
  \strid{mult_sym_rnat_r}
  =
  \strid{mult_sym_rnat_l}
  &
  \strid{mult_sym_lnat_r}
  =
  \strid{mult_sym_lnat_l}
  \\[8ex]
  \strid{eta_sym_rnat_l}
  =
  \strid{eta_sym_rnat_r}
  &
  \strid{eta_sym_lnat_l}
  =
  \strid{eta_sym_lnat_r}
\end{array}
\]
A \emph{commutative monoid} is a symmetric monoid such that the diagram
\[
\xymatrix{
  &M\otimes M\ar[dr]^{\mu}&\\
  M\otimes M\ar[ur]^{\gamma}\ar[rr]_{\mu}&&M
}
\]
commutes. Graphically,
\begin{equation}
  \label{eq:monoid_mult_comm}
  \strid{mult_comm}
  \qeq
  \strid{mult}
\end{equation}
A commutative monoid in a symmetric monoidal category is a commutative monoid
whose symmetry corresponds to the symmetry of the category:
$\gamma=\gamma_{M,M}$. In this case, the equations (\ref{eq:monoid-nat}) can
always be deduced from the naturality of the symmetry of the monoidal category.

A \emph{comonoid} $(M,\delta,\varepsilon)$ in $\mathcal{C}$ is an object $M$
together with two morphisms
\[
\delta:M\to M\otimes M
\qtand
\varepsilon:M\to I
\]
respectively drawn as
\[
\strid{comult_m}
\qqtand
\strid{counit_m}
\]
satisfying dual coherence diagrams. An similarly, the notions symmetric
comonoid, cocommutative comonoid and cocommutative comonoid can be defined by
duality.

\subsection{An equational theory of monoids}
\label{subsection:eq-th-monoid}
The definition of a monoid can be reformulated internally using the notion of
equational theory.
\begin{definition}
  \label{definition:e-t-monoid}
  The \emph{equational theory of monoids} $\eqth{M}$ has only one object $1$ and
  two generators $\mu:2\to 1$ and $\eta:0\to 1$ subject to the equations
  \[
  \mu\circ(\mu\otimes\id_1)
  =
  \mu\circ(\id_1\otimes\mu)
  \qtand
  \mu\circ(\eta\otimes\id_1)
  =
  \id_1
  =
  \mu\circ(\id_1\otimes\eta)
  \]
\end{definition}
We write $\mathcal{M}$ for the monoidal category generated by the equational
theory $\eqth{M}$. It can easily be seen that a monoid $M$ in a strict monoidal
category $\mathcal{C}$ is essentially the same as a functor from $\mathcal{M}$
to $\mathcal{C}$. More precisely,
\begin{property}
  The category $\Alg{\mathcal{M}}{\mathcal{C}}$ of algebras of the monoidal
  theory $\mathcal{M}$ in $\mathcal{C}$ is equivalent to the category of monoids
  in $\mathcal{C}$.
\end{property}
Similarly, all the algebraic structures introduced in this section can be
defined using algebraic theories.

\begin{remark}
  The presentations given here are not necessarily minimal. For example, in the
  theory of commutative monoids the equation on the right-hand side
  of~(\ref{eq:monoid_unit}) is derivable from the
  equation~(\ref{eq:monoid_mult_comm}), the equation on the left-hand side
  of~(\ref{eq:monoid_unit}) and one of the equations~(\ref{eq:monoid-nat}):
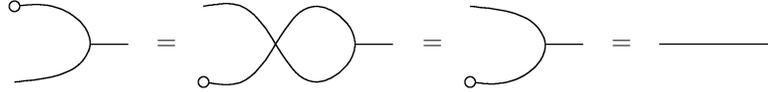
  \[
  \strid{mult_unit_r}
  =
  \strid{mult_unit_l_sym}
  =
  \strid{mult_unit_l}
  =
  \strid{mult_unit_c}
  \]
  A minimal presentation of this equational theory with three generators and
  seven equations is given in~\cite{massol:minimality}. However, not all the
  equational theories introduced in this paper have a known presentation which
  is proved to be minimal.
\end{remark}

\subsection{Bialgebras}
A \emph{bialgebra} $(B,\mu,\eta,\delta,\varepsilon,\gamma)$ in $\mathcal{C}$ is
an object $B$ together with four morphisms
\[
\begin{array}{r@{\quad:\quad}l}
  \mu&B\otimes B\to B\\
  \eta&I\to B\\
  \delta&B\to B\otimes B\\
  \varepsilon&B\to I\\
  \gamma&B\otimes B\to B\otimes B\\
\end{array}
\]
respectively drawn as
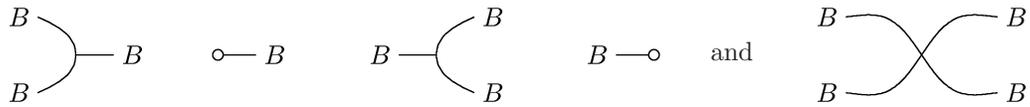
\[
\strid{mult_b}
\quad
\strid{unit_b}
\quad
\strid{comult_b}
\quad
\strid{counit_b}
\qtand
\strid{sym_b}
\]
such that $\gamma:B\otimes B\to B\otimes B$ is a symmetry for $B$,
$(B,\mu,\eta,\gamma)$ is a symmetric monoid and $(B,\delta,\varepsilon,\gamma)$
is a symmetric comonoid. Those two structures should be coherent, in the sense
that the diagrams
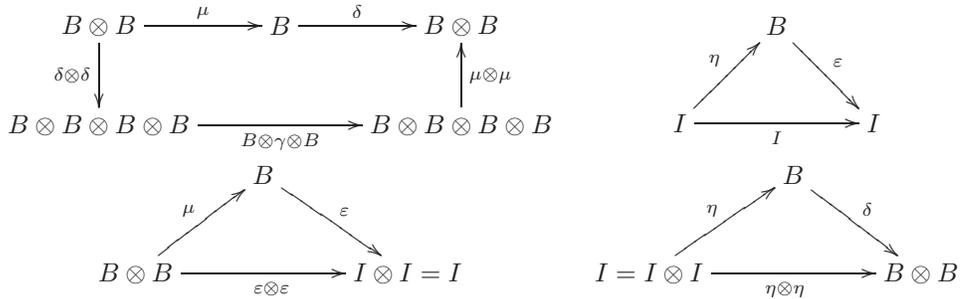
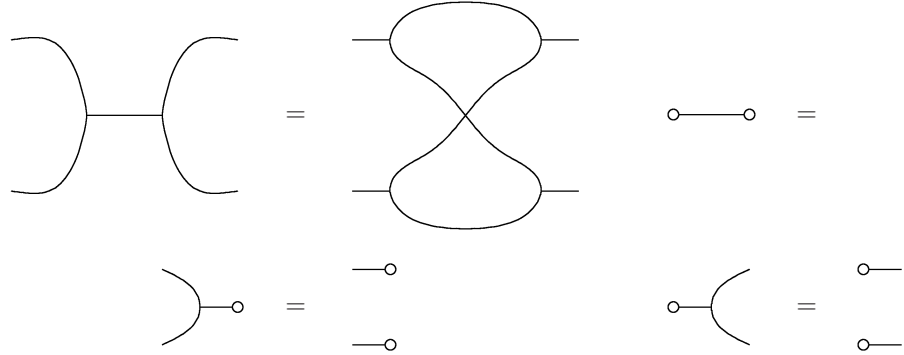
\[
\begin{array}{cc}
  \xymatrix{
    B\otimes B\ar[d]_{\delta\otimes\delta}\ar[r]^-{\mu}&B\ar[r]^-{\delta}&B\otimes B\\
    B\otimes B\otimes B\otimes B\ar[rr]_{B\otimes\gamma\otimes B}&&\ar[u]_{\mu\otimes\mu}B\otimes B\otimes B\otimes B\\
  }&
  \xymatrix{
    &B\ar[dr]^{\varepsilon}&\\
    I\ar[ur]^{\eta}\ar[rr]_{I}&&I
  }\\
  \xymatrix{
    &B\ar[dr]^{\varepsilon}&\\
    B\otimes B\ar[ur]^{\mu}\ar[rr]_{\varepsilon\otimes\varepsilon}&&I\otimes I=I
  }&
  \xymatrix{
    &B\ar[dr]^{\delta}&\\
    I=I\otimes I\ar[ur]^{\eta}\ar[rr]_{\eta\otimes\eta}&&B\otimes B
  }
\end{array}
\]
should commute. Graphically,
\[
\begin{array}{r@{\qeq}l@{\qquad}r@{\qeq}l}
  \strid{hopf_l}&\strid{hopf_r}
  &
  \strid{unit_counit}&
  \\
  \strid{counit_mult}&\strid{counit_x_counit}
  &
  \strid{comult_unit}&\strid{unit_x_unit}
\end{array}
\]

A morphism of bialgebras of $\mathcal{C}$
\[
f:(A,\mu_A,\eta_A,\delta_A,\varepsilon_A,\gamma_1)\to(B,\mu_B,\eta_B,\delta_B,\varepsilon_B,\gamma_B)
\]
is a morphism $f:A\to B$ of $\mathcal{C}$ preserving the structure of bialgebra,
that is
\[
(f\otimes f)\circ\mu_A=\mu_B\circ f
\qqcomma
f\circ\eta_A=\eta_B
\qqcomma
\text{\etc}
\]
The symmetric bialgebra is \emph{commutative} (\resp \emph{cocommutative}) when
the induced symmetric monoid $(B,\mu,\eta,\gamma)$ (\resp symmetric comonoid
$(B,\delta,\varepsilon,\gamma)$) is commutative (\resp cocommutative), and
\emph{bicommutative} when it is both commutative and cocommutative.

A \emph{qualitative bialgebra} is a bialgebra
$(B,\mu,\eta,\delta,\varepsilon,\gamma)$ such that the diagram
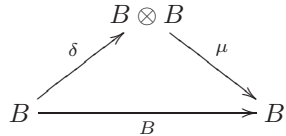
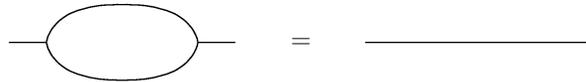
\[
\xymatrix{
  &B\otimes B\ar[dr]^{\mu}&\\
  B\ar[ur]^{\delta}\ar[rr]_{B}&&B
}
\]
commutes. Graphically,
\[
\strid{rel_l}
\qeq
\strid{rel_r}
\]

A bialgebra $B$ in a symmetric monoidal category is a bialgebra whose symmetry
morphism $\gamma$ corresponds with the symmetry of the category $\gamma_{B,B}$.

Similarly to what has been explained for monoids in
Section~\ref{subsection:eq-th-monoid}, an equational theory of bialgebras, \etc
can be defined. We write $\eqth{B}$ for the equational theory of bicommutative
bialgebras and $\eqth{R}$ for the equational theory of bicommutative qualitative
bialgebras.

\subsection{Dual objects}
An object $L$ of $\mathcal{C}$ is said to be \emph{left dual} to an object $R$
when there exists two morphism
\[
\eta:I\to R\otimes L
\qtand
\varepsilon:L\otimes R\to I
\]
called respectively the \emph{unit} and the \emph{counit} of the duality and
respectively pictured as
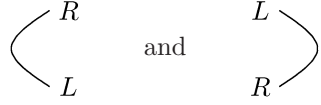
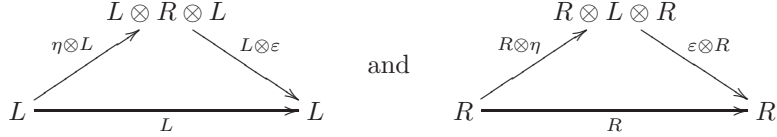
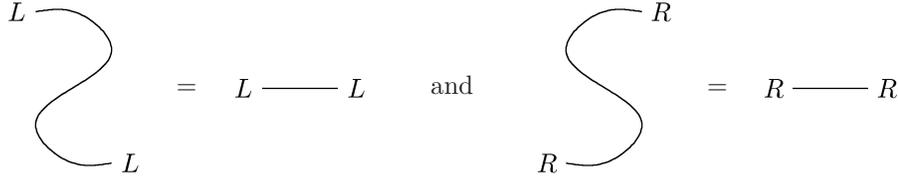
\[
\strid{adj_unit_lr}
\qtand
\strid{adj_counit_lr}
\]
making the diagrams
\[
\vxym{
  &L\otimes R\otimes L\ar[dr]^{L\otimes\varepsilon}&\\
  L\ar[ur]^{\eta\otimes L}\ar[rr]_L&&L\\
}
\qtand
\vxym{
  &R\otimes L\otimes R\ar[dr]^{\varepsilon\otimes R}&\\
  R\ar[ur]^{R\otimes\eta}\ar[rr]_{R}&&R\\
}
\]
commute. Graphically,
\[
\strid{zig_zag_l}
=
\strid{id_L}
\qtand
\strid{zig_zag_r}
=
\strid{id_R}
\]
We write $\eqth{D}$ for the equational theory associated to dual objects and
$\mathcal{D}$ for the generated monoidal category.

If $\mathcal{C}$ is category, two dual objects in the monoidal category
$\mathrm{End}(\mathcal{C})$ of endofunctors of $\mathcal{C}$, with tensor
product given on objects by composition of functors, are adjoint endofunctors of
$\mathcal{C}$. More generally, the theory of adjoint functors in a 2-category is
given in~\cite{street:free-adj}, the definition of $\eqth{D}$ is a
specialization of this construction to the case where there is only one 0-cell.

\section{A presentation of relations}
\label{section:presentation-rel}
We now introduce a presentation for the category $\Rel$ of finite ordinals and
relations. This result is mentioned in Examples~6 and~7
of~\cite{hyland-power:symmetric-monoidal-sketches} and is proved in three
different ways in~\cite{lafont:equational-reasoning-diagrams},
\cite{pirashvili:bialg-prop} and~\cite{lack:composing-props}. The proof we give
here has the advantage of being simple to check and can be extended to give a
presentation of the category of games and strategies, see
Section~\ref{subsection:walking-inno}.

\subsection{The simplicial category}
The simplicial category $\Delta$ is the strict monoidal category whose objects
are the finite ordinals and whose morphisms $f:\intset{m}\to\intset{n}$ are the
monotone functions from $\intset{m}$ to $\intset{n}$.

It has been known for a long time that this category is closely related to the
notion of monoid, see~\cite{maclane:cwm} or~\cite{lafont:boolean-circuits} for
example. This result can be formulated as follows:
\begin{property}
  \label{property:delta-presentation}
  The monoidal category $\Delta$ is presented by the equational theory of
  monoids $\eqth{M}$.
\end{property}
In this sense, the simplicial category $\Delta$ impersonates the notion of
monoid.

Dually, the monoidal category $\Delta^\op$, which is isomorphic to the category
of finite ordinals and (weakly) monotonic functions $f:\intset{m}\to\intset{n}$
such that $f(0)=0$, impersonates the notion of comonoid:
\begin{property}
  The monoidal category $\Delta^\op$ is presented by the equational theory of
  comonoids.
\end{property}


In the next Section, we show how to extend these results to the monoidal
category of multirelations.

\subsection{Multirelations}
A \emph{multirelation} $R$ between two sets $A$ and $B$ is a function from
$A\times B\to\N$. It can be equivalently be seen as a multiset whose elements
are in $A\times B$, or as a matrix over $\N$, or as a span
\[
\xymatrix@C=2ex@R=2ex{
  &\ar[dl]_sR\ar[dr]^t&\\
  A&&B\\
}
\]
in the category $\Set$ -- for the latest case, the multiset representation can
be recovered from the span by
\[
R(a,b)\qeq\left|\setof{e\in R\tq s(e)=a\tand t(e)=b}\right|
\]
for every element $(a,b)\in A\times B$. If $R_1:A\to B$ and $R_2:B\to C$ are two
multirelations, their composition is defined by
\[
R_2\circ R_1
\qeq
(a,c)\mapsto\sum_{b\in B}R_1(a,b)\times R_2(b,c)
\tdot
\]
Again, this corresponds to the usual composition of matrices if we see $R_1$ and
$R_2$ as matrices over $\N$, and as the span obtained by computing the pullback
\[
\xymatrix@C=2ex@R=2ex{
  &&\ar[dl]R_2\circ R_1\ar[dr]&&\\
  &\ar[dl]_{s_1}R_1\ar[dr]^{t_1}&&\ar[dl]_{s_2}R_2\ar[dr]^{t_2}&\\
  A&&B&&C\\
}
\]
if we see $R_1$ and $R_2$ as spans in $\Set$.

The cardinal $\card{R}$ of a multirelation $R:A\to B$ is defined by
\[
|R|\qeq\sum_{(a,b)\in A\times B}R(a,b)
\tdot
\]

We write $\FMR$ for the monoidal theory of multirelations: its objects are
finite ordinals and morphisms are multirelations between them. It is a strict
symmetric monoidal category with the tensor product $\otimes$ defined on two
morphisms $R_1:\intset{m_1}\to\intset{n_1}$ and
$R_2:\intset{m_2}\to\intset{n_2}$ by
\[
R_1\otimes R_2=R_1\cup R_2:\intset{m_1}+\intset{m_2}\to\intset{n_1}+\intset{n_2}
\]
and the morphisms
\[
R^\gamma_{\intset{m},\intset{n}}=
(\intset{m}\times\intset{n})\cup(\intset{n}\times\intset{m}):
\intset{m}+\intset{n}\to\intset{n}+\intset{m}
\]
as symmetry. In particular, the following multirelations are morphisms in
$\FMR$:
\[
\begin{array}{r@{\quad:\quad}l}
  R^\mu=(i,j)\mapsto 1&\intset{2}\to\intset{1}\\
  R^\eta=(i,j)\mapsto 1&\intset{0}\to\intset{1}\\
  R^\delta=(i,j)\mapsto 1&\intset{1}\to\intset{2}\\
  R^\varepsilon=(i,j)\mapsto 1&\intset{1}\to\intset{0}\\
  R^\gamma=(i,j)\mapsto
  \begin{cases}
    0&\text{if $i=j$,}\\
    1&\text{otherwise.}
  \end{cases}&\intset{2}\to\intset{2}
\end{array}
\]

We now show that multirelations are presented by the equational theory
$\eqth{B}$ bicommutative bialgebras. We write $\mathcal{B}$ for the monoidal
category generated by $\eqth{B}$.

\begin{lemma}
  \label{property:fmr-bialg}
  In $\FMR$, $(1,R^\mu,R^\eta,R^\delta,R^\varepsilon)$ is a bicommutative
  bialgebra.
\end{lemma}

For every morphism $\phi:m\to n$ in $\mathcal{B}$, where $m>0$, we define a
morphism $S\phi:m+1\to n$ by
\[
S\phi\qeq \phi\circ(\gamma\otimes \id_{m-1})
\]
We introduce the following notation which is defined inductively by
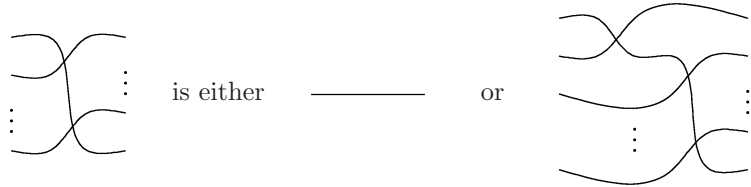
\[
\strid{gsym}
\quad\text{is either}\quad
\strid{gsym_id}
\qtor
\strid{gsym_sym}
\]
These morphisms are called \emph{stairs}: a stair is therefore either $\id_1$ or
$S\phi'$ where $\phi'$ is a stairs. The \emph{length} of a stairs is defined as
$0$ if its of the first form and the length of the stairs plus one if it is of
the second form.

We define the following notion of \emph{canonical form} inductively: $\phi$ is
either
\begin{equation}
  \label{eq:bialg-nf-eta}
  \strid{bialg_nf_eta}
\end{equation}
or there exists a canonical form $\phi'$ such that $\phi$ is either
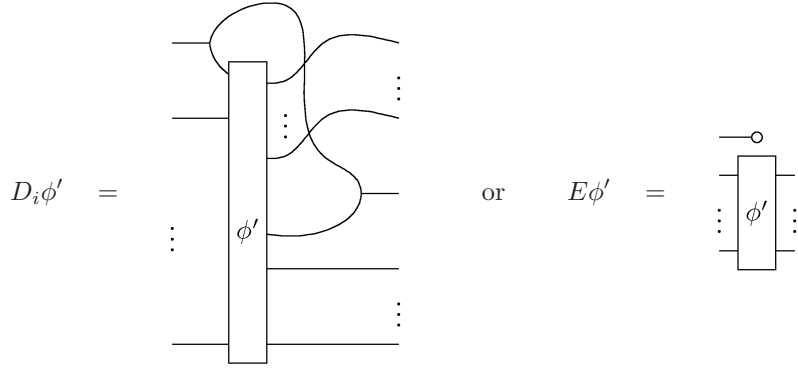
\[
D_i\phi'\qeq\strid{bialg_nf_mu}
\qqtor
E\phi'\qeq\strid{bialg_nf_eps}
\]
In the latter case we write respectively $\phi$ as $D_i\phi'$ (where the index
$i$ is the length of the stairs) or as $E\phi'$.

Showing that identities are equal to canonical forms require the slightly more
general following lemma.
\begin{lemma}
  \label{lemma:bialg-nf-id}
  Any morphism $f=\eta\otimes\cdots\otimes\eta\otimes\id_n:m\to m$ is equal to a
  canonical form.
\end{lemma}
\begin{proof}
  By induction on $n$. The result is immediate when $n=0$. Otherwise, we have
  the equalities of Figure~\ref{fig:bialg-nf-id-proof} in Appendix which show
  that $f$ is equal to a morphism of the form $D_i(E\phi)$, where $\phi$ is
  equal to a canonical form by induction hypothesis.
\end{proof}

\begin{lemma}
  \label{lemma:bialg-nf-comm}
  For every morphism $\phi:m\to n$, where $m>0$, for all indices $i$ and $j$
  such that $0\leq i\leq n$ and $0\leq j\leq n$, we have
  \[
  D_j(D_i\phi)\qeq D_i(D_j\phi)
  \]
\end{lemma}
\begin{proof}
  The proof is done by examining separately the cases $j<i$, $i\neq j$ and $j>i$
  and showing the result for each case using in particular the derivable
  equalities shown in Figure~\ref{fig:bialg-stairs-yb} in Appendix.
\end{proof}

\noindent
From this we deduce that
\begin{lemma}
  Every multirelation $R:m\to n$ is represented by a canonical form and two
  canonical forms representing $R$ are equal.
\end{lemma}
\begin{proof}
  This is proved by induction on $m$ and on the cardinal $\card{R}$ of $R$.
  \begin{enumerate}
  \item If $m=0$ then $R$ is represented by a unique normal form which is of the
    form (\ref{eq:bialg-nf-eta}).
  \item If $m>0$ and for every $j<n$, $R(0,j)=0$ then $R$ is of the form
    $R=R^\varepsilon\otimes R'$ and $R$ is necessarily represented by a
    canonical form $E\phi'$ where $\phi'$ is a canonical form representing
    $R':(m-1)\to n$, which exists by induction hypothesis.
  \item Otherwise, $R$ is necessarily represented by a canonical form of the
    form $D_k\phi'$, where $k$ is such that $R(0,k)>0$ and $\phi'$ is a
    canonical form represented by the relation $R':m\to n$ defined by
    \[
    R'(i,j)=
    \begin{cases}
      R(i,j)-1&\text{if $i=0$ and $j=k$,}\\
      R(i,j)&\text{else.}\\
    \end{cases}
    \]
    and such a canonical form exists by induction hypothesis.
  \end{enumerate}
  By Lemma~\ref{lemma:bialg-nf-comm}, two canonical forms $\phi_1$ and $\phi_2$
  representing $R$, obtained by choosing different values for $k$ in case~3
  during the construction of the canonical form are equal.
\end{proof}

\begin{lemma}
  \label{lemma:bialg-nf-init}
  Every morphism $f:m\to n$ in $\mathcal{B}$ is equal to a canonical form.
\end{lemma}
\begin{proof}
  The proof is done by induction on the size $\size{f}$ of $f$.
  \begin{itemize}
  \item If $\size{f}=0$ then $m=n$ and $f=\id_m$ which is equal to a canonical
    form by Lemma~\ref{lemma:bialg-nf-id}.
  \item If $\size{f}>0$ then $f$ is of the form $f=h\circ g$ where $\size{h}=1$
    and $\size{g}=\size{f}-1$. By induction hypothesis, $g$ is equal to a
    canonical form $\phi$. Since $h$ is of size $1$, it is of the form
    $h=\id_{m_2}\circ h'\circ\id_{m_1}$ where $h$ is either $\mu$, $\eta$,
    $\delta$, $\varepsilon$ or $\gamma$. We show the result by case
    analysis. For the lack of space, we only detail the case where
    $h'=\mu$. There are four cases to handle which are shown in
    Figures~\ref{fig:bialg-nf-init-proof1}, \ref{fig:bialg-nf-init-proof2},
    \ref{fig:bialg-nf-init-proof3} and~\ref{fig:bialg-nf-init-proof4}.
  \end{itemize}
\end{proof}

\begin{theorem}
  The category $\FMR$ of multirelations is presented by the equational theory
  $\eqth{B}$ of bicommutative bialgebras.
\end{theorem}

\subsection{Relations}
The monoidal category $\Rel$ has finite ordinals as objects and relations as
morphisms. This category can be obtained from $\FMR$ by quotienting the
morphisms by the equivalence relation $\sim$ on multirelations defined as
follows. Two multirelations $R_1,R_2:m\to n$ are such that $R_1\sim R_2$
whenever
\begin{equation}
\label{eq:multirel-rel}
\forall i<m, \forall j<n,\quad R_1(i,j)\neq 0\tiff R_2(i,j)\neq 0
\end{equation}
This induces a full monoidal functor $F$ from $\FMR$ to $\Rel$. We still write
$R^\mu$, $R^\eta$, $R^\delta$, $R^\varepsilon$ and $R^\gamma$ for the images by
this functor of the corresponding multirelations. We denote $\mathcal{R}$ for
the monoidal category generated by the equational theory $\eqth{R}$ of
qualitative bicommutative bialgebras.

\begin{lemma}
  \label{lemma:rel-nf-rel}
  For every morphism $\phi:m\to n$ in $\mathcal{R}$, where $m>0$, for every
  index $i$ such that $0\leq i\leq n$, we have
  \[
  D_i(D_i\phi)\qeq D_i\phi
  \]
\end{lemma}
\begin{proof}
  See Figure~\ref{fig:rel-nf-rel-proof} in Appendix.
\end{proof}

\noindent
From this Lemma, we deduce that:
\begin{theorem}
  The category $\Rel$ of relations is presented by the equational theory
  $\eqth{R}$ of qualitative bicommutative bialgebras.
\end{theorem}
\begin{proof}
  Since $\Rel$ can be obtained from $\FMR$ by quotienting morphisms, by
  Lemma~\ref{property:fmr-bialg},
  $(1,R^\mu,R^\eta,R^\delta,R^\varepsilon,R^\gamma)$ is still a bialgebra in
  $\Rel$ and moreover it satisfies the additional equation making it a
  qualitative bialgebra. Therefore $\Rel$ is a model of the equational theory
  $\eqth{R}$. Moreover, $\eqth{R}$ is a complete axiomatization of $\Rel$. In
  order to show this, we use the same notion of canonical form as in the
  previous Section: we have to show that two canonical forms representing the
  same relation are equal. This amounts to check that two canonical forms
  representing two multirelations $R_1$ and $R_2$, which are equivalent by the
  relation (\ref{eq:multirel-rel}), are equal. This can easily be done using
  Lemmas~\ref{lemma:bialg-nf-comm} and~\ref{lemma:rel-nf-rel}.
\end{proof}

\section{A game semantics for first-order propositional logic}
\label{section:games-strategies}

\subsection{First-order propositional logic}
Suppose that we are given a fixed first-order language~$\mathcal{L}$, that is
\begin{enumerate}
\item a set of proposition symbols~$P,Q,\ldots$ with given arities,
\item a set of function symbols~$f,g,\ldots$ with given arities,
\item a set of first-order variables~$x,y,\ldots$.
\end{enumerate}
\emph{Terms}~$t$ and \emph{formulas}~$A$ are respectively generated by the
following grammars:
\[
t\qgramdef x\gramor f(t,\ldots,t)
\qquad\qquad
A\qgramdef P(t,\ldots,t)\gramor\qforall{x}{A}\gramor\qexists{x}{A}
\]
We suppose that application of propositions and functions always respect
arities. Moreover, we suppose here that there are proposition and function
symbols of any arity (this is needed for the definability result of
Proposition~\ref{prop:definability}).
Formulas are considered modulo renaming of variables. Substitution $A[t/x]$ of a
free variable $x$ by a term $t$ in a formula $A$ is defined as usual, avoiding
capture of variables.
We consider the logic associated to these formulas, where proofs are generated
by the following inference rules:
\[
\begin{array}{c@{\qquad}c}
  \inferrule{A[t/x]\vdash B}{\qforall x A\vdash B}{\lrule{$\forall$-L}}
  &
  \inferrule{A\vdash B}{A\vdash \qforall x B}{\lrule{$\forall$-R}}
  \\
  &
  \text{(with $x$ not free in $A$)}
  \\[2ex]
  \inferrule{A\vdash B}{\qexists x A\vdash B}{\lrule{$\exists$-L}}
  &
  \inferrule{A\vdash B[t/x]}{A\vdash \qexists x B}{\lrule{$\exists$-R}}
  \\
  \text{(with $x$ not free in $B$)}
  &
  \\[2ex]
  \inferrule{\null}{P(t_1,\ldots,t_n)\vdash P(t_1,\ldots,t_n)}{\lrule{Ax}}
  &
  \inferrule{A\vdash B\\B\vdash C}{A\vdash C}{\lrule{Cut}}
  \\[2ex]
\end{array}
\]

\subsection{Games and strategies}
\label{subsection:games-strategies}
\begin{definition}
A \emph{game} $A=(\moves{A},\lambda_A,\leq_A)$ consists of a set of moves
$\moves{A}$, a polarization function $\lambda_A:\moves{A}\to\{-1,+1\}$ which to
every move $m$ associates its polarity, and a partial order $\leq_A$ on moves
such that every move $m\in\moves{A}$ defines a finite downward closed set
\[
m\!\downarrow\qeq\setof{n\in\moves{A}\tq n\leq_A m}
\tdot
\]
\end{definition}
A move $m$ is said to be a Proponent move when $\lambda_A(m)=+1$ and an Opponent
move else.

Suppose that $A$ and $B$ are two games. Their tensor product $A\otimes B$ is
defined by
\[
\moves{A\otimes B}=\moves{A}\uplus\moves{B}
\tcomma\quad
\lambda_{A\otimes B}=\lambda_A+\lambda_B
\qtand
\leq_{A\otimes B}=\leq_A\cup\leq_B
\tdot
\]
The opposite game $A^*$ of the game $A$ is defined by
\[
A^*=(\moves{A},-\lambda_A,\leq_A)
\tdot
\]
Finally, the arrow game $A\llimp B$ is defined by
\[
A\llimp B\qeq A^*\otimes B
\tdot
\]
A game $A$ is \emph{filiform} when the associated partial order is total.

Two partial orders $\leq$ and $\leq'$ on a set $M$ are \emph{compatible} when
their relational union $\leq\cup\leq'$ is still an order (\ie is acyclic).

\begin{definition}
  A \emph{strategy} $\sigma$ on a game $A$ is a partial order $\leq_\sigma$ on
  the moves of $A$ which is compatible with the order of the game and is
  moreover such that for every moves $m,n\in\moves{A}$,
  \begin{equation}
    \label{eq:strategy-polarisation}
    m<_\sigma n\qtimpl\lambda_A(m)=-1\tand\lambda_A(n)=+1
    \tdot
  \end{equation}
\end{definition}
The \emph{size} $\size{A}$ of a game $A$ is the cardinal of $\moves{A}$ and the
\emph{size} $\size{\sigma}$ of a strategy $\sigma:A$ is the cardinal of the set
\[
\setof{(m,n)\in\moves{A}\times\moves{A}\tq m<_\sigma n}
\tdot
\]

If $\sigma:A\llimp B$ and $\tau:B\llimp C$ are two strategies, their composite
$\tau\circ\sigma:A\llimp C$ is the partial order $\leq_{\tau\circ\sigma}$ on the
moves of $A\llimp C$, defined as the restriction to the set of moves of $A\llimp
C$ of the transitive closure of the union~$\leq_\sigma\cup\leq_\tau$ of the
partial orders~$\leq_\sigma$ and~$\leq_\tau$ considered as relations. The
identity strategy $\id_A:A\llimp A$ on a game $A$ is the strategy such that for
every move $m$ of $A$ we have $m_L\leq_{\id_A}m_R$ if $\lambda(m)=-1$ and
$m_R\leq_{\id_A}m_L$ if $\lambda(m)=+1$ when $m_L$ (\resp $m_R$) is the instance
of the move $m$ in the left-hand side (\resp right-hand side) copy of $A$. It
can easily be checked that for every strategy $\sigma:A\to B$ we have
$\id_B\circ\sigma=\sigma=\sigma\circ\id_A$.

Since the composition of strategies is defined in the category of relations, we
still have to check that the composite of two strategies $\sigma$ and $\tau$ is
actually a strategy. Preservation of the polarization condition
(\ref{eq:strategy-polarisation}) by composition is easily checked. However,
proving that the relation $\leq_{\tau\circ\sigma}$ corresponding to the
composite strategy is acyclic is more difficult: a direct proof of this property
is combinatorial and a bit lengthy. For now, we define the category~$\Games$ as
the smallest category whose objects are filiform games, whose morphisms between
two games $A$ and $B$ contain the strategies on the game $A\llimp B$ and is
moreover closed under composition. We will deduce in
Corollary~\ref{corol:composition} that strategies are in fact the only morphisms
of this category from our presentation of the category.

If $A$ and $B$ are two games, the game $A\before{}B$ (to be read $A$ \emph{before}
$B$) is the game defined by
\[
M_{A\before{}B}=M_A\uplus M_B
\qcomma
\lambda_{A\before{}B}=\lambda_A+\lambda_B
\]
and $\leq_{A\before{}B}$ is the transitive closure of the relation
\[
\leq_A\cup\leq_B\cup\;\setof{(a,b)\tq a\in M_A\tand b\in M_B}
\]
This operation is extended as a bifunctor on strategies as follows. If
$\sigma:A\to B$ and $\tau:C\to D$ are two strategies, the strategy
$\sigma\before{}\tau:A\before{}C\to B\before{}D$ is defined as the transitive
closure of the relation
\[
\leq_{\sigma\before{}\tau}
\qeq
\leq_\sigma\cup\leq_\tau
\]
This bifunctor induces a monoidal structure $(\Games,\before{},I)$ on the
category $\Games$, where $I$ denotes the empty game.

We write $O$ for a game with only one Opponent move and $P$ for a game with only
one Proponent move. It can be easily remarked that filiform games $A$ are
generated by the following grammar
\[
A\qqgramdef I\gramor O\before{}A\gramor P\before{}A
\]
A game $X_1\before{}\cdots\before{}X_n\before{}I$ where the $X_i$ are either $O$
or $P$ is represented graphically as
\[
\xymatrix@R=1ex{
  X_1\\
  \vdots\\
  X_n\\
}
\]
A strategy $\sigma:A\to B$ is represented graphically by drawing a line from a
move $m$ to a move $n$ whenever $m\leq_\sigma n$. For example, the strategy
$\mu^P:P\before{}P\to P$
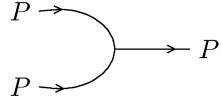
\[
\strid{mult_P}
\]
is the strategy on $(O\before{}O)\otimes P$ in which both Opponent move of the
left-hand game justify the Proponent move of the right-hand game. When a move
does not justify (or is not justified by) any other move, we draw a line ended
by a small circle. For example, the strategy $\varepsilon^P:P\to I$ drawn as
\[
\strid{counit_P}
\]
is the unique strategy from $P$ to $I$.

\begin{figure}[h!]
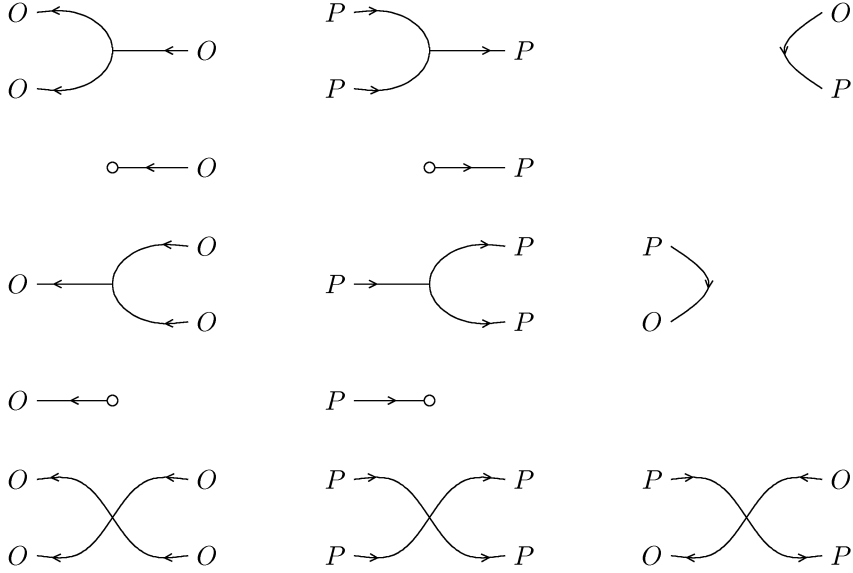

  \[
  \begin{array}{r@{\qcolon}l@{\quad}r@{\qcolon}l@{\quad}rcl}
    \mu^O&O\otimes O\to O&\mu^P&P\otimes P\to P&\eta^{OP}&\colon&I\to O\otimes P\\
    \eta^O&I\to O&\eta^P&I\to P\\
    \delta^O&O\to O\otimes O&\delta^P&P\to P\otimes P&\varepsilon^{OP}&\colon&P\otimes O\to I\\
    \varepsilon^O&O\to I&\varepsilon^P&P\to I\\
    \gamma^O&O\otimes O\to O\otimes O&\gamma^P&P\otimes P\to P\otimes P&\gamma^{OP}&\colon&P\otimes O\to O\otimes P\\
  \end{array}
  \]
  respectively drawn as
  \[
  \begin{array}{c@{\qquad}c@{\qquad}c}
    \strid{mult_O}&\strid{mult_P}&\strid{unit_OP}\\
    \strid{unit_O}&\strid{unit_P}\\
    \strid{comult_O}&\strid{comult_P}&\strid{counit_OP}\\
    \strid{counit_O}&\strid{counit_P}\\
    \strid{sym_O}&\strid{sym_P}&\strid{sym_OP}\\
  \end{array}
  \]
  \caption{Generators of the strategies.}
  \label{fig:inno-gen}
\end{figure}

With these conventions, we introduce notations for some morphisms which are
depicted in Figure~\ref{fig:inno-gen} (perhaps a bit confusingly, the tensor
product $\otimes$ on this figure is the $\before{}$ tensor).

\subsection{A game semantics for proofs}
A formula $A$ is interpreted as a game $\intp{A}$ by
\[
\intp{P}=I
\qquad
\intp{\qforall x A}=O\before{}\intp{A}
\qquad
\intp{\qexists x A}=P\before{}\intp{A}
\]

A proof $\pi:A\vdash B$ is interpreted as the strategy $\sigma:A\llimp B$. The
corresponding partial order $\leq_\sigma$ is defined as follows. For every
Proponent move $P$ interpreting a quantifier which is introduced by a rule
\[
\inferrule{A[t/x]\vdash B}{\qforall x A\vdash B}{\lrule{$\forall$-L}}
\qqtor
\inferrule{A\vdash B[t/x]}{A\vdash \qexists x B}{\lrule{$\exists$-R}}
\]
every Opponent move $O$ interpreting an universal quantification $\forall x$ on
the right-hand side of a sequent, or an existential quantification $\exists x$
on the left-hand side of a sequent, is such that $O\leq_\sigma P$ whenever the
variable $x$ is free in the term $t$. The partial order interpreting a proof
$\pi$ can easily be shown to be a strategy.

For example, a proof
\[
\inferrule{
\inferrule{
\inferrule{
\inferrule{\null}
{P\vdash Q[t/z]}{\lrule{Ax}}
}
{P\vdash\qexists z Q}{\lrule{$\exists$-R}}
}
{\qexists y P\vdash\qexists z Q}{\lrule{$\exists$-L}}
}
{\qexists x{\qexists y P}\vdash\qexists z Q}{\lrule{$\exists$-L}}
\]
is interpreted respectively by the strategies
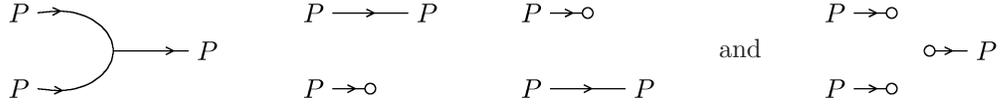
\[
\strid{strat_ex_xy}
\quad
\strid{strat_ex_x}
\quad
\strid{strat_ex_y}
\qtand
\strid{strat_ex_}
\]
when the free variables of $t$ are $\{x,y\}$, $\{x\}$, $\{y\}$ and $\emptyset$.

The following Proposition shows that our game semantics contains only definable
strategies.
\begin{proposition}
  \label{prop:definability}
  For every strategy $\sigma:A\to B$ in $\Games$, there exists two propositions
  $P$ and $Q$ such that $A=\intp{\sqcap_1\ldots\sqcap_k P}$,
  $B=\intp{\sqcap_1\ldots\sqcap_l Q}$ and there exists a proof
  $\pi:\sqcap_1\ldots\sqcap_k P\vdash \sqcap_1'\ldots\sqcap_l' Q$ such that
  $\intp{\pi}=\sigma$, where $\sqcap_i$ and $\sqcap_i'$ is either $\forall$ or
  $\exists$.
\end{proposition}

\subsection{An equational theory of strategies}
\label{subsection:walking-inno}

\begin{definition}
  \label{definition:innocent-strategies}
  The \emph{equational theory of strategies} is the equational theory $\eqth{G}$
  with two types $O$ and $P$ and 13 generators depicted in
  Figure~\ref{fig:inno-gen} such that
  \begin{itemize}
  \item $(O,\mu^O,\eta^O,\delta^O,\varepsilon^O,\gamma^O)$ is a bicommutative
    qualitative bialgebra,
  \item the Proponent structure is adjoint to the Opponent structure in the
    sense that the equations of Figure~\ref{fig:PO-adj} hold.
  \end{itemize}
\end{definition}
We write $\mathcal{G}$ for the monoidal category generated by $\eqth{G}$.

\begin{remark}
  The generators $\mu^P$, $\eta^P$, $\delta^P$, $\varepsilon^P$, $\gamma^P$ and
  $\gamma^{OP}$ are superfluous in this presentation. However, removing them
  would seriously complicate the proofs.
\end{remark}

\begin{lemma}
  With the notations of~\ref{definition:innocent-strategies}, we have:
  \begin{itemize}
  \item $(P,\mu^P,\eta^P,\delta^P,\varepsilon^P,\gamma^P)$ is a qualitative
    bicommutative bialgebra,
  \item the Yang-Baxter equalities
    \[
    \strid{yang_baxter_xyz_r}
    \qeq
    \strid{yang_baxter_xyz_l}
    \]
    hold whenever $(X,Y,Z)$ is either $(O,O,O)$, $(P,O,O)$, $(P,P,O)$ or
    $(P,P,P)$,
  \item the equalities
    \[
      \strid{mult_sym_rnat_P_l}
      =
      \strid{mult_sym_rnat_P_r}
    \]
    and
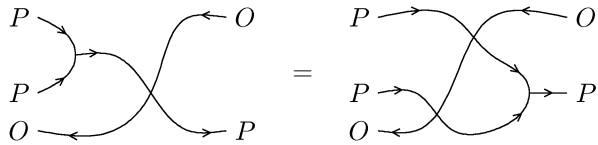
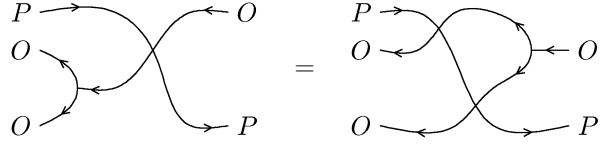
    \[
    \strid{mult_sym_lnat_O_l}
    =
    \strid{mult_sym_lnat_O_r}
    \]
    hold (and dually for comultiplications),
  \item the equalities
    \[
    \strid{eta_sym_rnat_P_l}
    =
    \strid{eta_sym_rnat_P_r}
    \]
    and
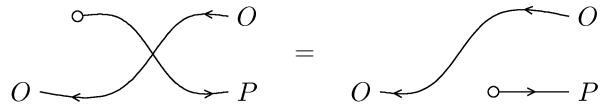
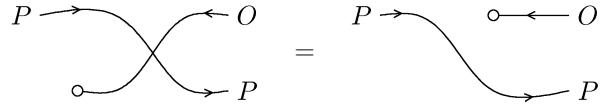
    \[
    \strid{eta_sym_lnat_O_l}
    =
    \strid{eta_sym_lnat_O_r}
    \]
    hold (and dually for counits),
  \item the equalities
    \[
    \strid{adj_counit_O_r}
    =
    \strid{adj_counit_O_l}
    \]
    and
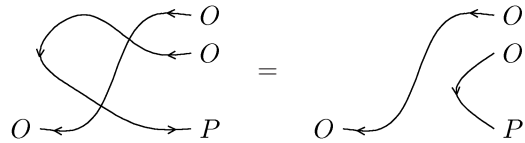
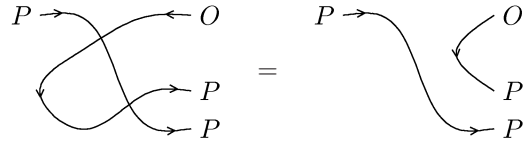
    \[
    \strid{adj_counit_P_r}
    =
    \strid{adj_counit_P_l}
    \]
    hold (and dually for the counit of duality).
  \end{itemize}
\end{lemma}


\begin{property}
  In the category $\Games$ with the monoidal structure induced by $\before{}$,
  the games $O$ and $P$ together with the morphisms introduced at the end of
  Section~\ref{subsection:games-strategies} induce a strategy structure in the
  sense of Definition~\ref{definition:innocent-strategies}.
\end{property}

We extend the proofs of Section~\ref{section:presentation-rel} to show that
$\eqth{G}$ is a presentation of the category $\Games$.

\emph{Stairs} are defined inductively by
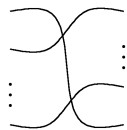
\[
\strid{gsym}
\]
is either
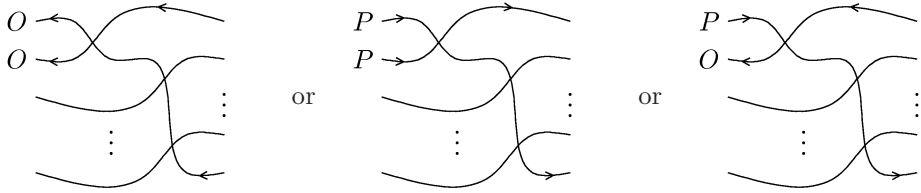
\[
\strid{gsym_id_O}
\qtor
\strid{gsym_id_P}
\]
or
\[
\strid{gsym_sym_O}
\tor
\strid{gsym_sym_P}
\tor
\strid{gsym_sym_OP}
\]
A \emph{canonical form} is either of the form $\phi=\psi\circ\theta$
\begin{equation}
  \label{eq:inno-nf-eta}
  \text{$\phi$ is}
  \qquad
  \strid{theta_psi}
\end{equation}
where a morphism of the form $\theta$ is defined inductively by
\begin{equation}
  \label{eq:inno-nf-theta}
  \text{$\theta$ is either void or}
  \qquad
  \strid{theta_nf_adj}
\end{equation}
where $\theta'$ is of the form (\ref{eq:inno-nf-theta}), and $\psi$ is defined
inductively by
\begin{equation}
  \label{eq:inno-nf-psi}
  \text{$\psi$ is either void or}
  \qquad
  \strid{psi_nf_eta}
  \qtor
  \strid{psi_nf_id}
\end{equation}
where $X$ is either $P$ or $O$ and $\psi'$ is of the form
(\ref{eq:inno-nf-psi}), or there exists a canonical form $\phi'$ such that
$\phi$ is either
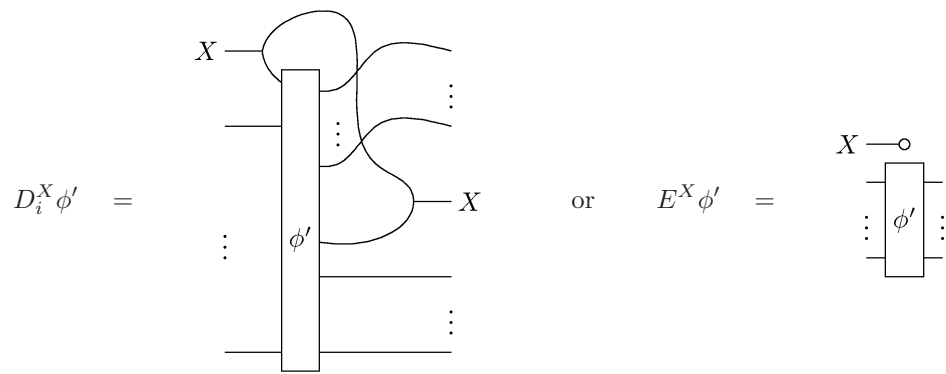
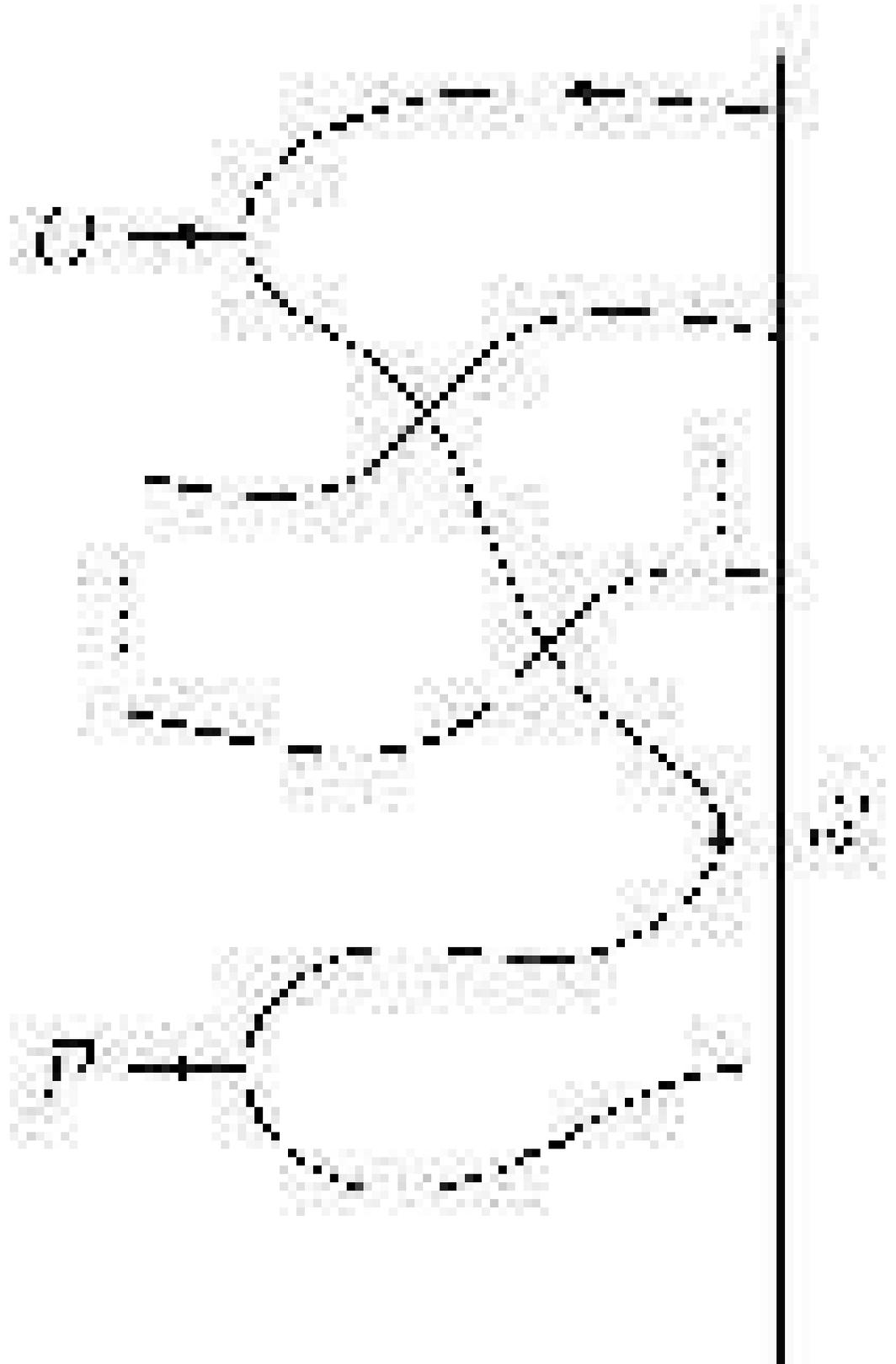
\[
D_i^X\phi'\qeq\strid{nf_mu}
\qqtor
E^X\phi'\qeq\strid{nf_eps}
\]
or
\[
A_i\phi'\qeq\strid{nf_adj}
\]
where $X$ is either $P$ or $O$. In the latter case, we write respectively $\phi$
as $D_i^X\phi'$ (where $i$ is the length of the stairs), or as $E^X\phi'$ or as
$A_i\phi'$ (where $i$ is the length of the stairs).

\begin{lemma}
  \label{lemma:inno-nf-comm}
  For any morphism $\phi$, we have
  \[
  \begin{array}{r@{\qeq}l@{\qquad\qquad}r@{\qeq}l}
    D_j^X(D_i^X\phi)&D_i^X(D_j^X\phi)&A_i(A_j\phi)&A_j(A_i\phi)\\
    D_i^X(D_i^X\phi)&D_i^X\phi&D_i^O(A_j\phi)&A_j(D_i^O\phi)\\
  \end{array}
  \]
  whenever both members of the equalities are defined, where $X$ is either $P$
  or $O$.
\end{lemma}

\begin{lemma}
  Every strategy $\sigma:A\to B$ is represented by a canonical form and two
  canonical forms representing the same strategy are equal.
\end{lemma}
\begin{proof}
  This is proved by induction on the respective sizes $\size{A}$ and
  $\size{\sigma}$ of $A$ and $\sigma$.
  \begin{enumerate}
  \item If $\size{A}=0$ then $\sigma$ is necessarily represented by canonical
    form of the form (\ref{eq:inno-nf-eta}), which is unique.
  \item If $A=X\before{}A'$, where $X$ is either $P$ or $O$ and
    $\moves{X}=\{m\}$, and for every move $n\in\moves{A\llimp B}$ we have
    $m\not<_\sigma n$ then $\sigma$ is necessarily represented by a canonical
    form $E^X\phi'$ where $\phi'$ is a canonical form representing the
    restriction of $\sigma$ to $A'\llimp B$.
  \item Otherwise, $A$ is of the form $A=X\before{}A'$, where $X$ is either $P$
    or $O$ and $\moves{X}=\{m\}$. A canonical form $\phi$ of $\sigma$ is
    necessarily of one of the two following forms.
    \begin{itemize}
    \item $\phi=D^X_i\phi'$ where $n$ is the $i$-th move of $B$ and is such that
      $m<_\sigma n$, and $\phi'$ is a canonical form representing either the
      strategy $\sigma$ or the strategy $\sigma'$ which is the same strategy as
      $\sigma$ excepting that $m\not<_{\sigma'}n$ -- for the construction part
      of the lemma we obviously chose the second possibility in order for the
      induction to work.
    \item $\phi=A_i\phi'$ where $n$ is the $i$-th move of $A$ and is such that
      $m<_\sigma n$, and $\phi'$ is a canonical form representing the strategy
      $\sigma'$ which is the same strategy as $\sigma$ excepting
      that $m\not<_{\sigma'}n$.
    \end{itemize}
    By Lemma~\ref{lemma:inno-nf-comm}, two canonical forms $\phi_1$ and $\phi_2$
    representing $\sigma$, obtained by choosing different values for $n$ in case
    3 are equal.
  \end{enumerate}
\end{proof}

\begin{lemma}
  Every morphism $f:A\to B$ of $\mathcal{G}$ is equal to a canonical form.
\end{lemma}
\begin{proof}
  The proof is similar to the proof of Lemma~\ref{lemma:bialg-nf-init}.
\end{proof}

\begin{theorem}
  The category $\Games{}$ is presented by the equational theory $\eqth{G}$.
\end{theorem}

As a direct consequence of this Theorem, we deduce that
\begin{corollary}
  \label{corol:composition}
  The composite of two strategies is a strategy.
\end{corollary}
In particular, acyclicity is preserved by composition.

\section{Conclusion}
We have constructed a game semantics for the fragment of first-order
propositional logic without connectives and given a presentation of the category
$\Games$ of games and definable strategies. Our methodology has proved very
useful to ensure that the composition of strategies was well-defined.

\bigskip

We consider this work much more as a starting point to bridge semantics and
algebra than as a final result. The methodology presented here seem to be very
general and many tracks remain to be explored.

First, we would like to extend the presentation to a game semantics for richer
logic systems like first-order propositional logic with conjunction and
disjunction. Whilst we do not expect many technical complications, this case is
much more difficult to grasp and manipulate since a presentation of such a
semantics would be a 4-polygraph (one dimension is added since games would be
trees instead of lines) and corresponding diagrams now live in a 3-dimensional
space.

It would be interesting to know whether it is possible to orient the equalities
in the presentations in order to obtain strongly normalizing rewriting systems
for the algebraic structures described in the paper. Such rewriting systems are
given in~\cite{lafont:boolean-circuits} -- for monoids and commutative monoids
for example -- but finding a strongly normalizing rewriting system presenting
the theory of bialgebras is still an open problem.

Finally, many of the proofs given here are repetitive and we believe that many
of them could be (at least partly) automated or mechanically checked. However,
finding a good representation of diagrams, in order for a program to be able to
manipulate them, is a difficult task that we should address in subsequent works.

\subsection*{Acknowledgements}
I would like to thank my PhD supervisor Paul-André Melliès as well as Yves
Lafont, Martin Hyland and Albert Burroni for the lively discussion we had, in
which I learned so many things.


\newpage
\bibliographystyle{plainnat}
\bibliography{these}

\newpage
\appendix
\section{Figures}
\begin{figure}[ht]
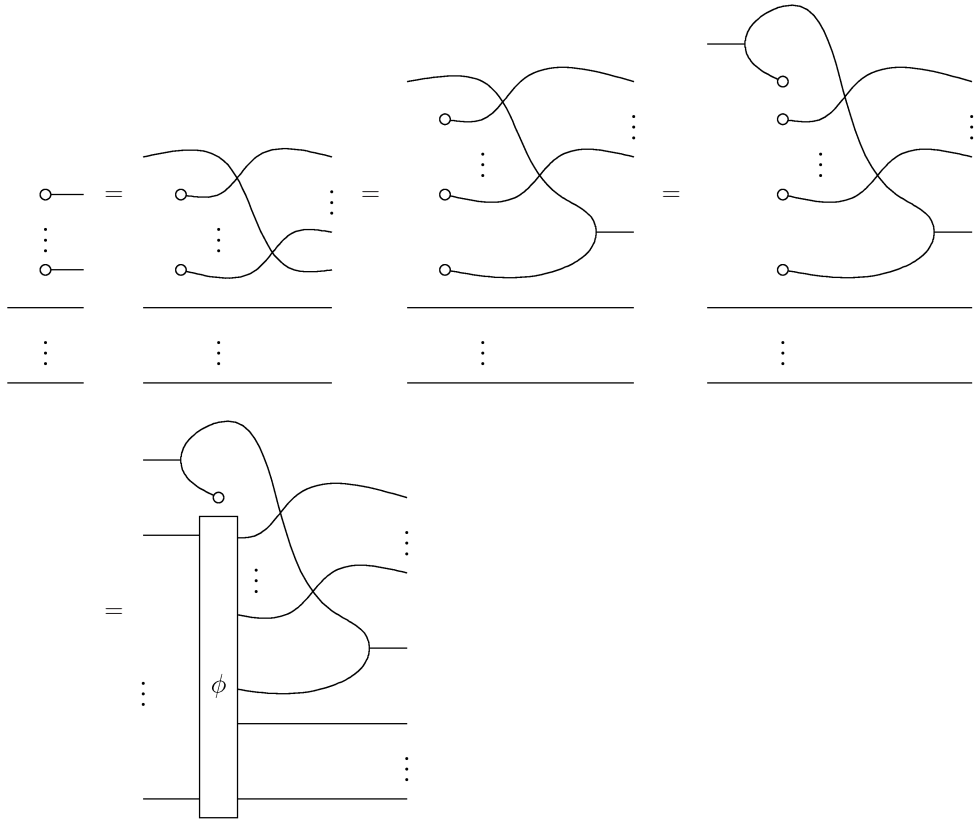

  \[
  \begin{array}{r@{=}l}
    \strid{eta_id_1}
    &
    \strid{eta_id_2}
    =
    \strid{eta_id_3}
    =
    \strid{eta_id_4}
    \\
    &
    \strid{eta_id_5}
  \end{array}
  \]
  \caption{Induction step in proof of Lemma~\ref{lemma:bialg-nf-id}.}
  \label{fig:bialg-nf-id-proof}
\end{figure}
\begin{figure}[ht]
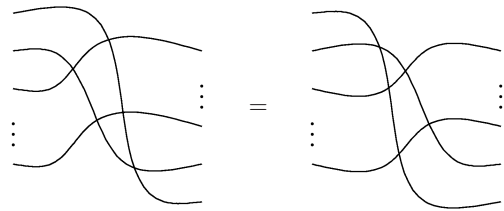

  \[
  \begin{array}{r@{\qeq}l}
    \strid{gsym_yb_l}
    &
    \strid{gsym_yb_r}
  \end{array}
  \]
  \caption{A generalization of the Yang-Baxter equality to stairs.}
  \label{fig:bialg-stairs-yb}
\end{figure}
\begin{figure}[ht]
  $g$ is of the form $D_i\phi'$ and $f$ is equal to
  \[
  \strid{bialg_nf_mu_mu_case1_1}
  \qeq
  \strid{bialg_nf_mu_mu_case1_2}
  \]
  which is of the form $D_j\phi$ where $\phi$ is equal to a canonical form
  by induction hypothesis.
  \caption{First case in proof of Lemma~\ref{lemma:bialg-nf-init}.}
  \label{fig:bialg-nf-init-proof1}
\end{figure}
\begin{figure}[ht]
  $g$ is of the form $D_i\phi'$ and and $f$ is equal to
  \[
  \begin{array}{cccc}
    &\strid{bialg_nf_mu_mu_case2_1}
    &=&
    \strid{bialg_nf_mu_mu_case2_2}
    \\
    =&
    \strid{bialg_nf_mu_mu_case2_3}
    &=&
    \strid{bialg_nf_mu_mu_case2_4}
    \\
  \end{array}
  \]
  which is of the form $D_j(D_k\phi)$ where $\phi$ is equal to a canonical
  form by induction hypothesis.
  \caption{Second case in proof of Lemma~\ref{lemma:bialg-nf-init}.}
  \label{fig:bialg-nf-init-proof2}
\end{figure}
\begin{figure}[ht]
  $g$ is of the form $D_i\phi'$ and and $f$ is equal to
  \[
  \strid{bialg_nf_mu_mu_case3_1}
  \]
  which is of the form $D_j\phi$ where $\phi$ is equal to a canonical form
  by induction hypothesis.
  \caption{Third case in proof of Lemma~\ref{lemma:bialg-nf-init}.}
  \label{fig:bialg-nf-init-proof3}
\end{figure}
\begin{figure}[ht]
  $g$ is of the form $E\phi'$ and and $f$ is equal to
  \[
  \strid{bialg_nf_eta_mu_case1_1}
  \]
  which is of the form $E\phi$ where $\phi$ is equal to a canonical form by
  induction hypothesis.
  \caption{Fourth case in proof of Lemma~\ref{lemma:bialg-nf-init}.}
  \label{fig:bialg-nf-init-proof4}
\end{figure}

\begin{figure}[ht]
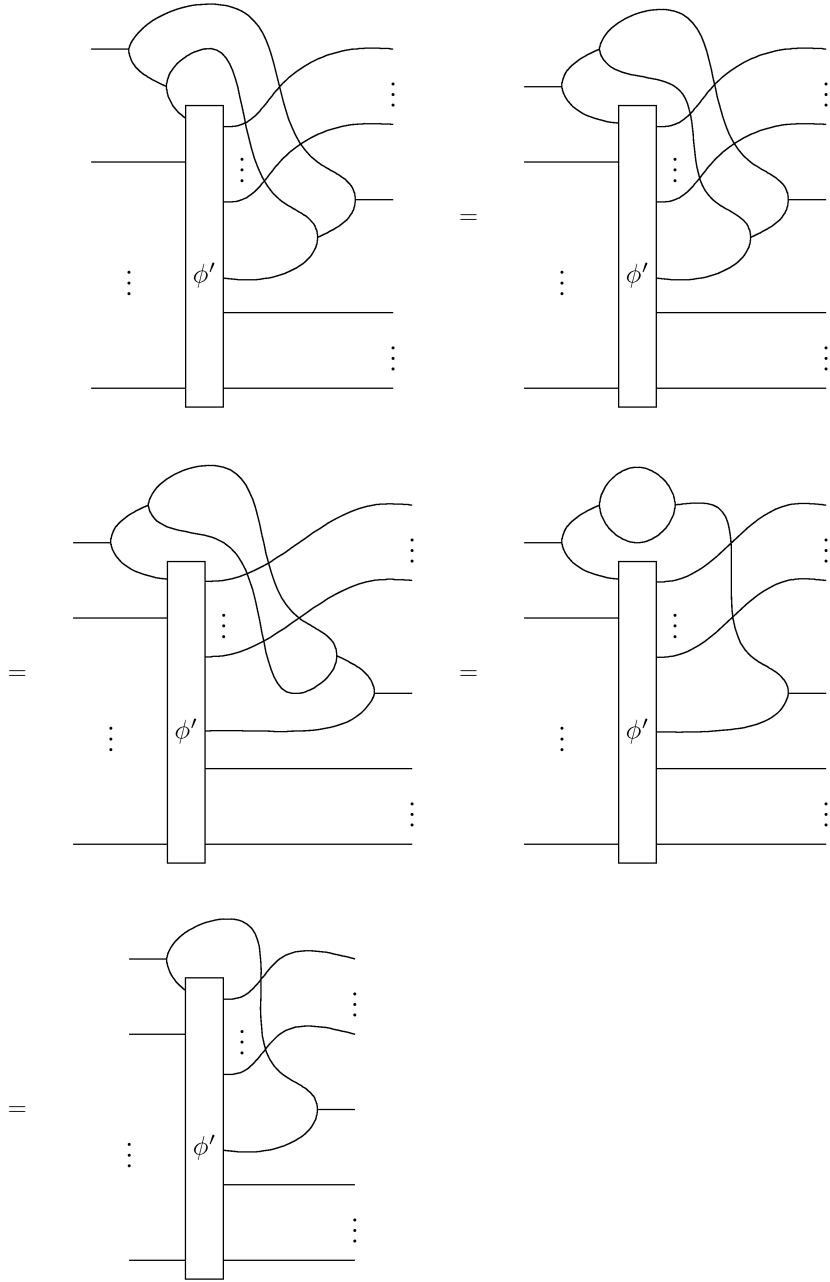

  \centering
  \[
  \begin{array}{cccc}
    &\strid{rel_nf_mu_mu_1}
    &=&
    \strid{rel_nf_mu_mu_2}
    \\
    =&
    \strid{rel_nf_mu_mu_3}
    &=&
    \strid{rel_nf_mu_mu_4}
    \\
    =&
    \strid{bialg_nf_mu}
  \end{array}
  \]
  \caption{Proof of Lemma~\ref{lemma:rel-nf-rel}.}
  \label{fig:rel-nf-rel-proof}
\end{figure}
\begin{figure}[ht]
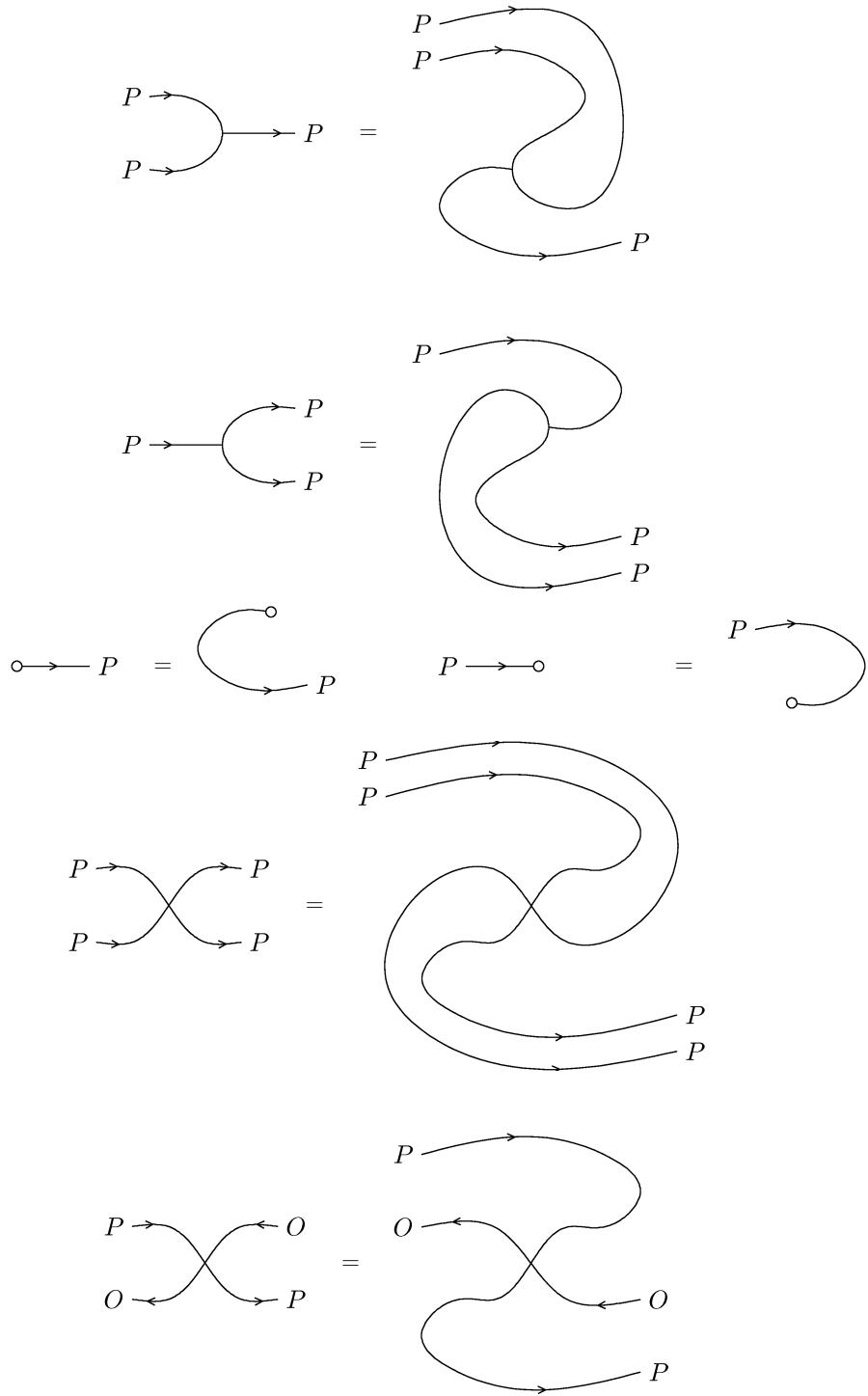

  \[
  \begin{array}{c}
    \strid{mult_P}
    =
    \strid{comult_O_adj}
    \\[4ex]
    \strid{comult_P}
    =
    \strid{mult_O_adj}
    \\[4ex]
    \strid{unit_P}
    =
    \strid{counit_O_adj}
    \qquad
    \strid{counit_P}
    =
    \strid{unit_P_adj}
    \\[4ex]
    \strid{sym_P}
    =
    \strid{sym_O_adj}
    \\[18ex]
    \strid{sym_OP}
    =
    \strid{sym_O_adj_OP}
    \\
  \end{array}
  \]
  \caption{Proponent is left dual to Opponent.}
  \label{fig:PO-adj}
\end{figure}

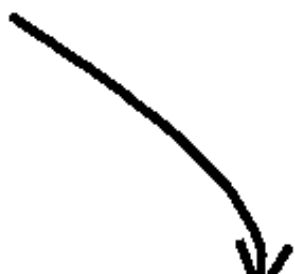
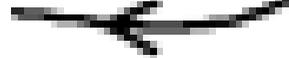
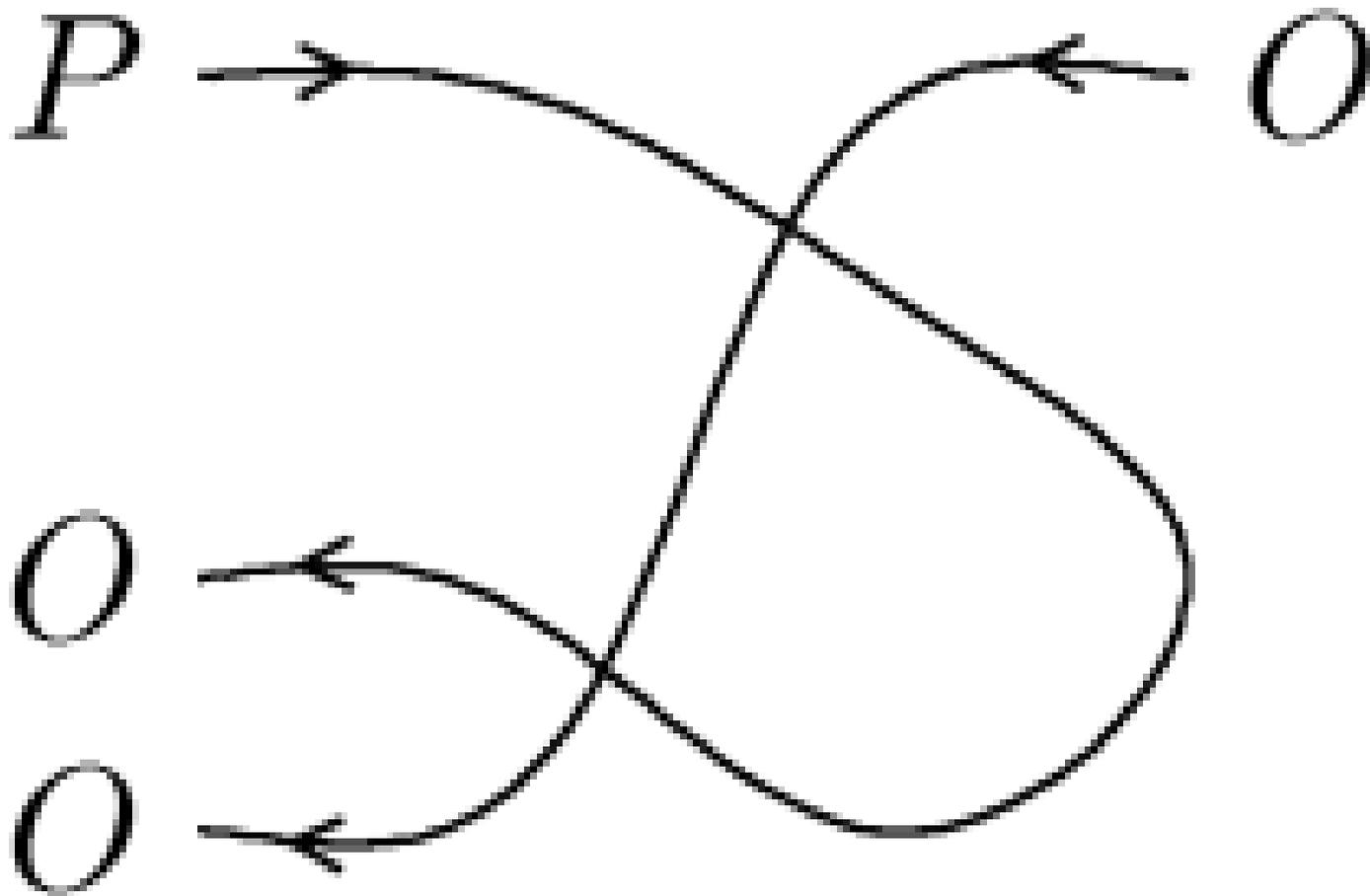
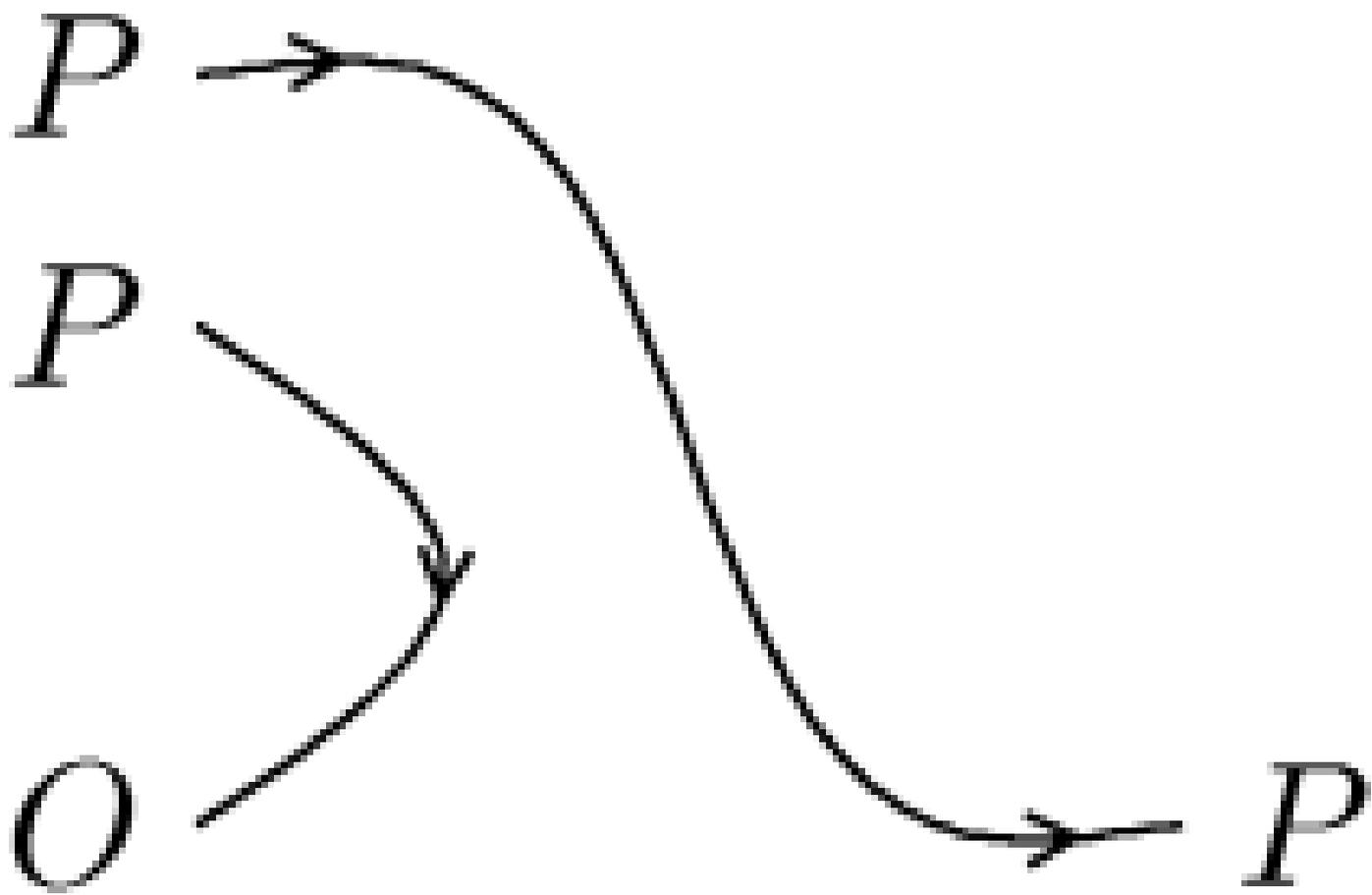
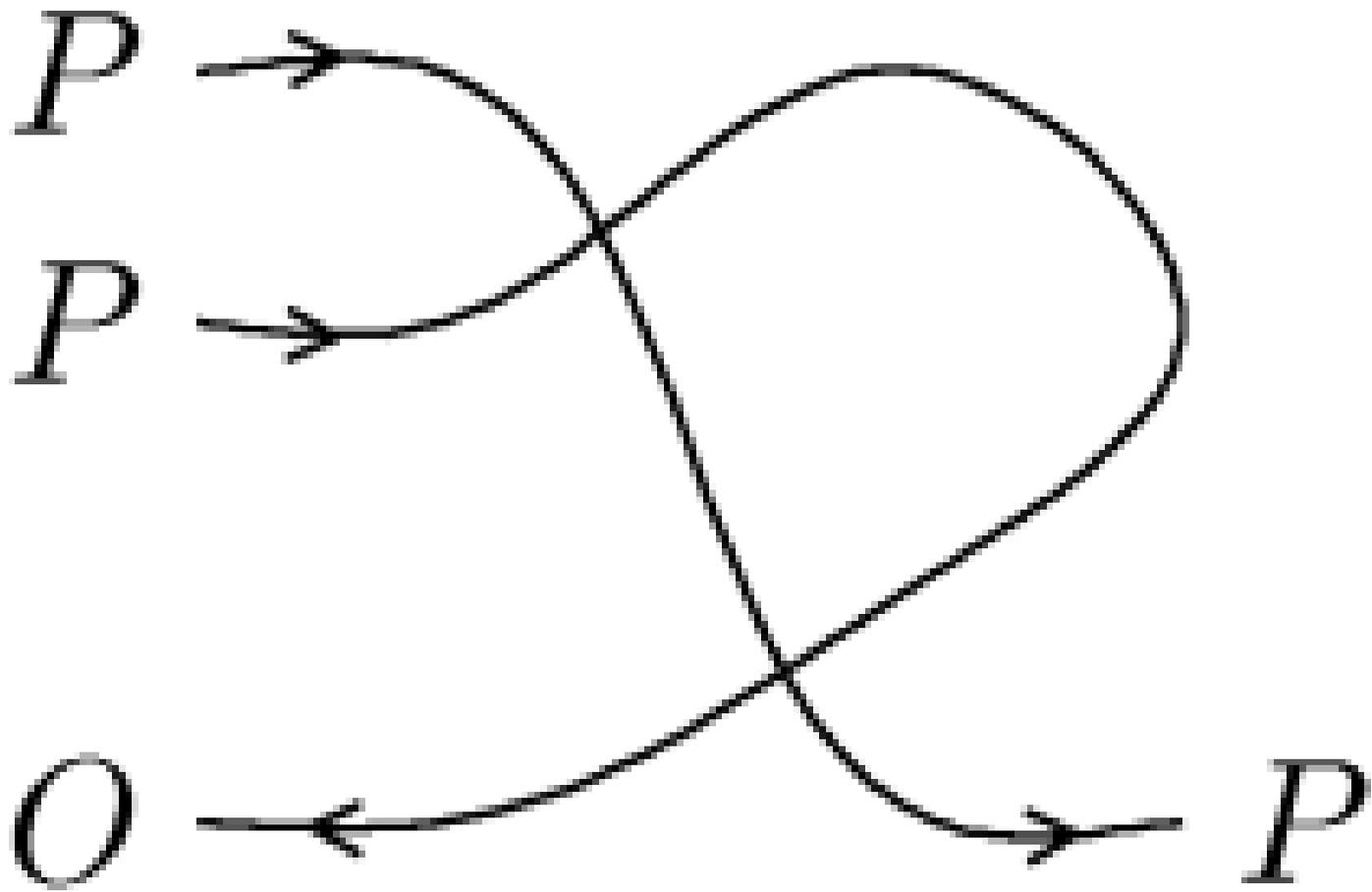
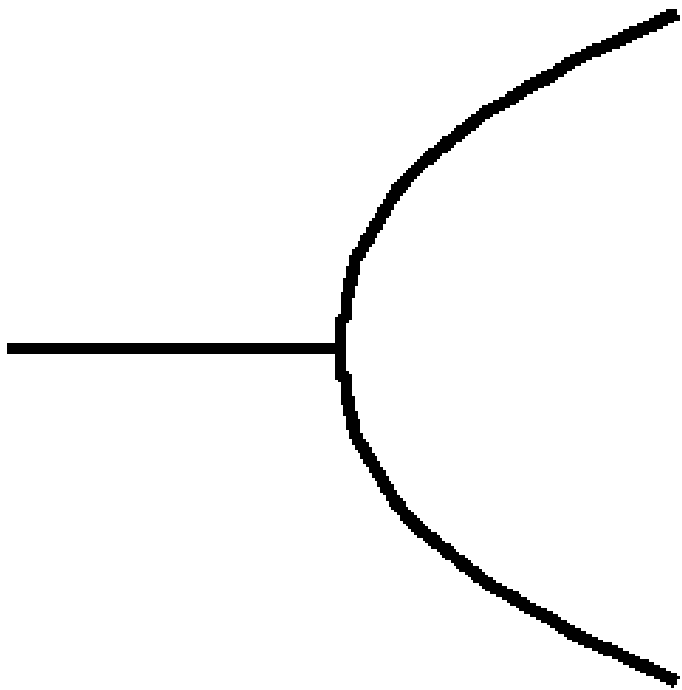
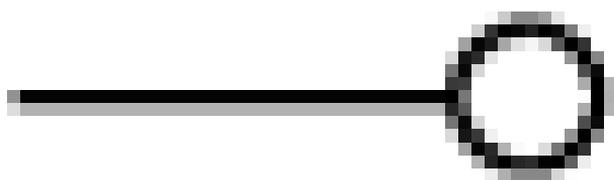
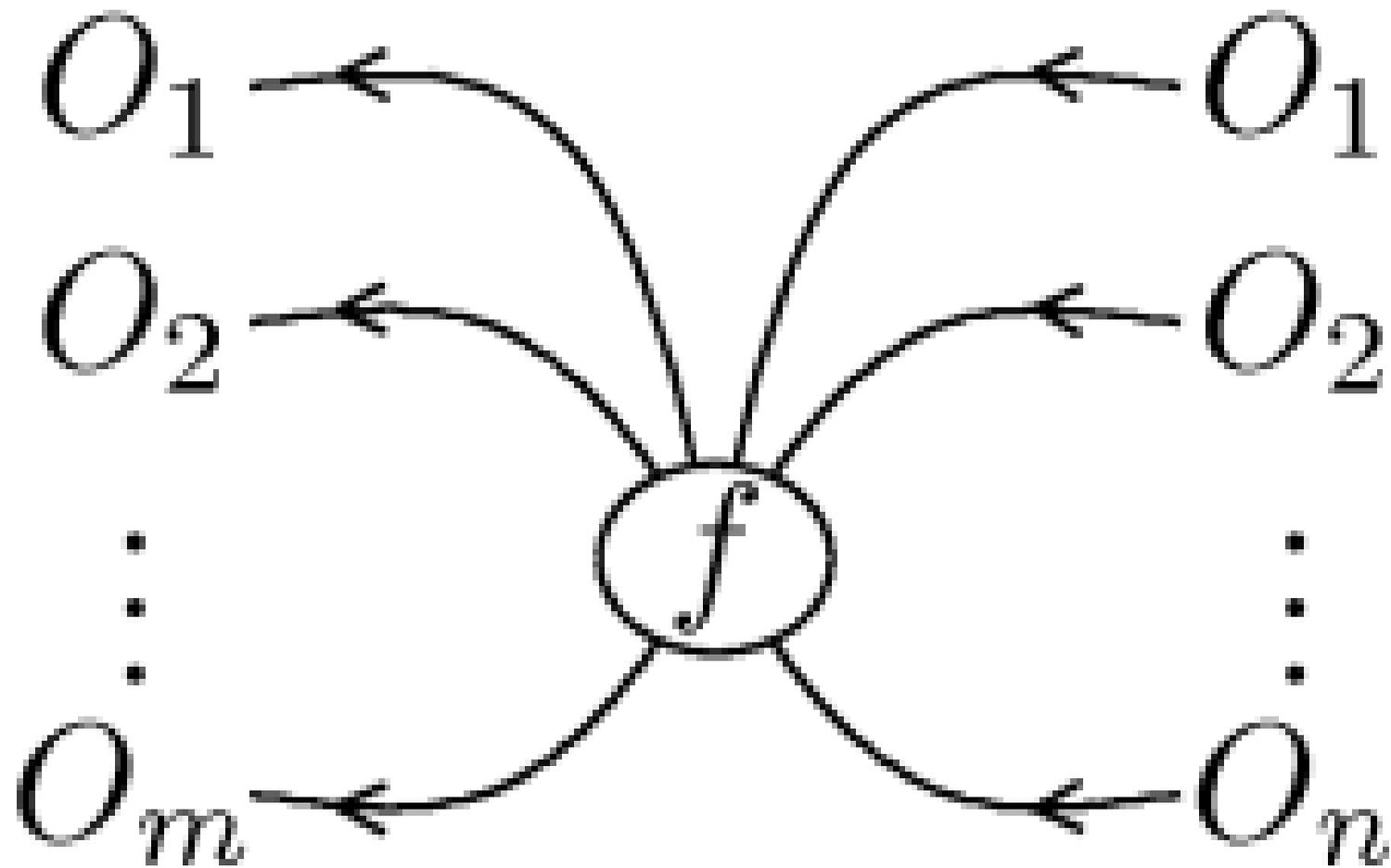
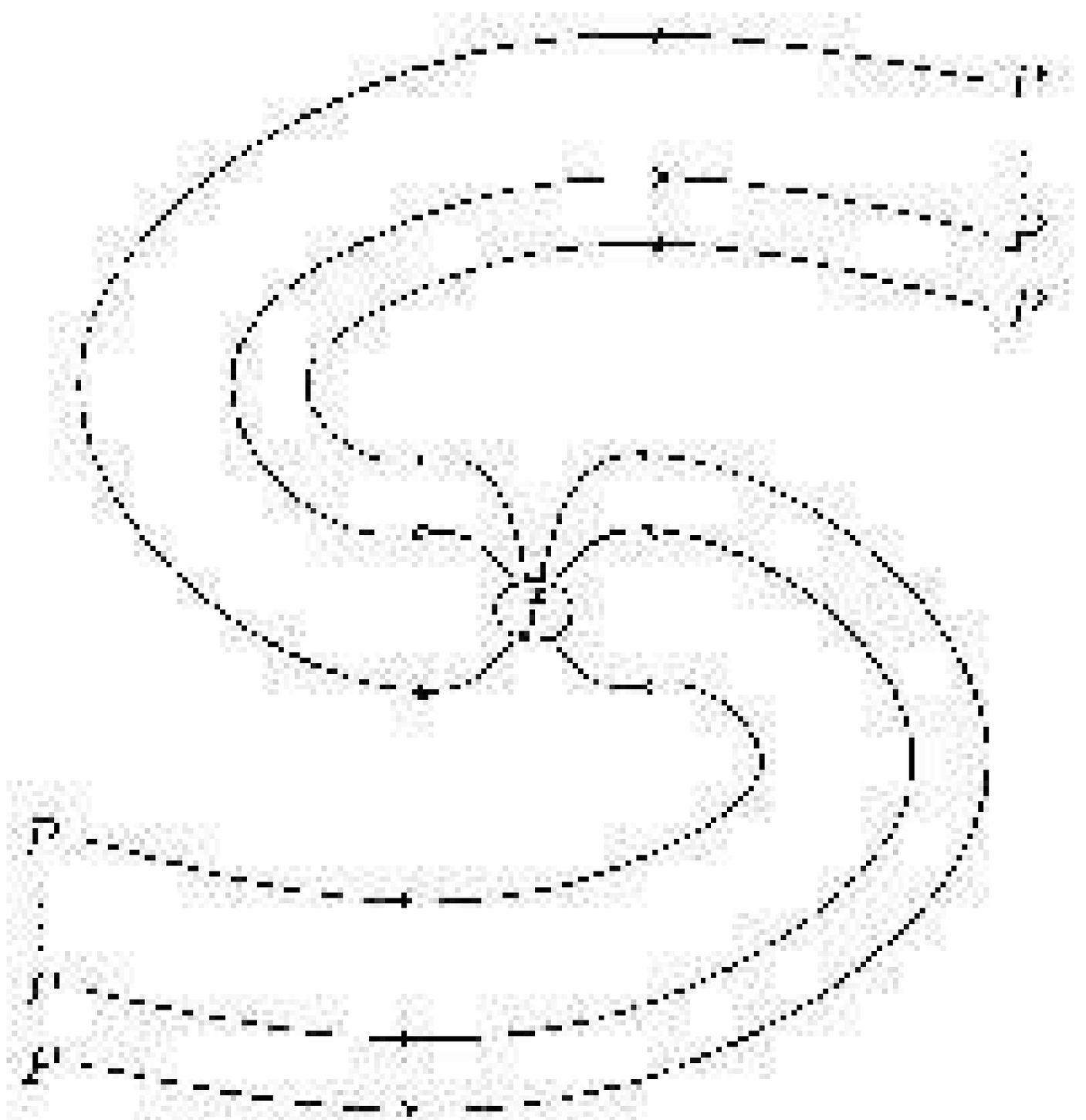
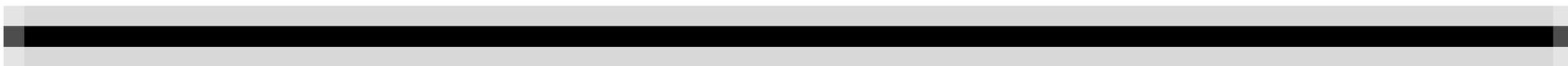
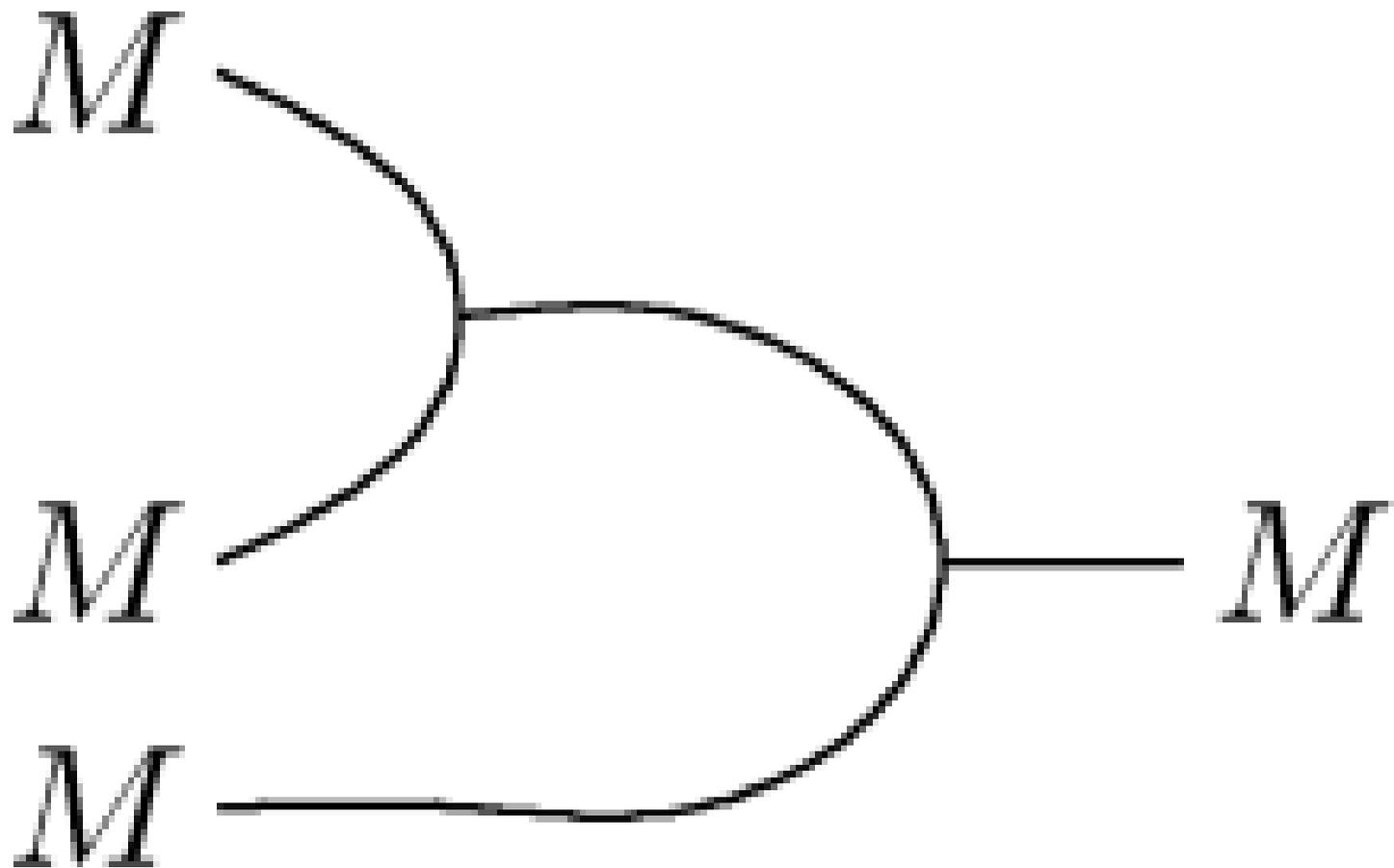
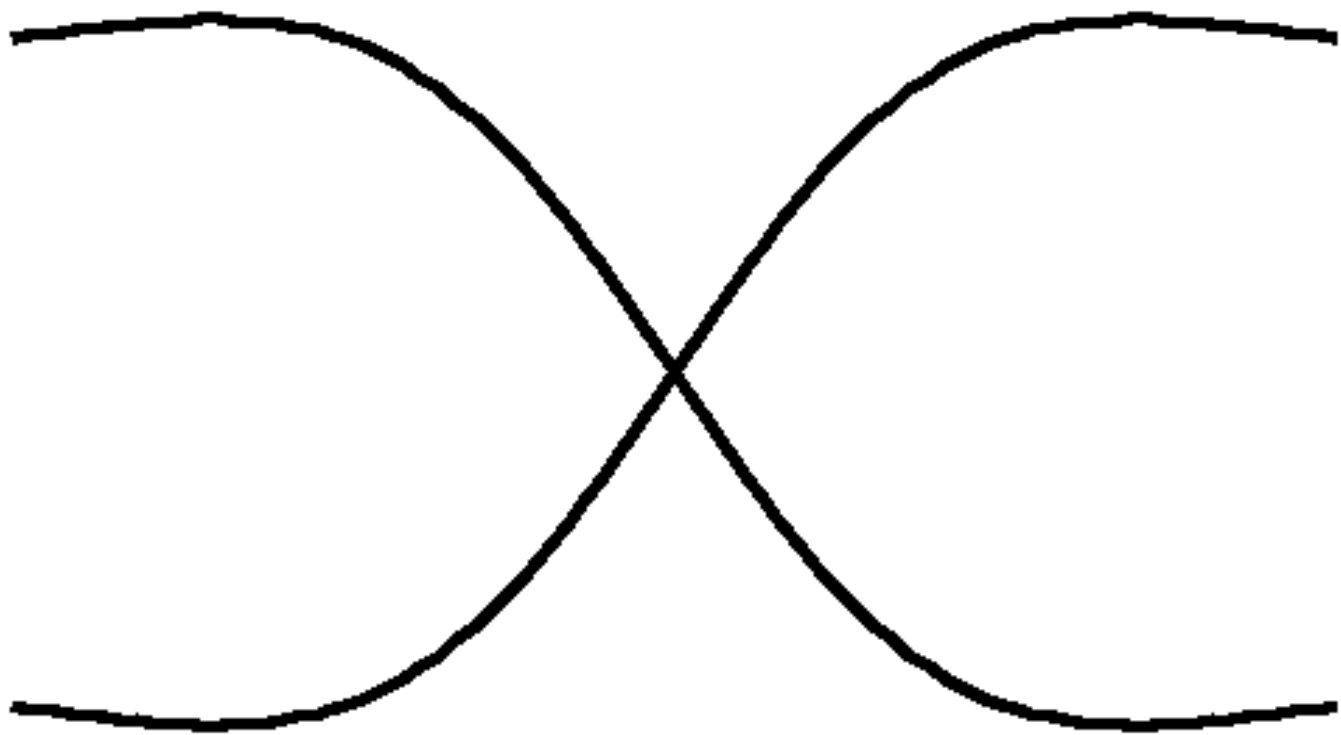
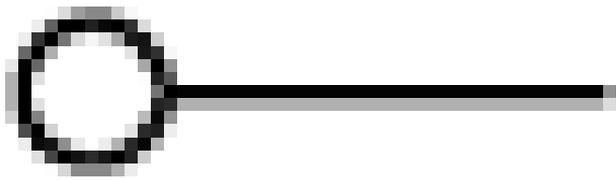
\end{document}